# Evolutionary constraints on the complexity of genetic regulatory networks allow predictions of the total number of genetic interactions

**Authors:** Adrian I. Campos-González[1,2] and Julio A. Freyre-González[1,]*.

**Affiliations:**

[1]Regulatory Systems Biology Research Group, Laboratory of Systems and Synthetic Biology, and
[2]Undergraduate Program in Genomic Sciences, Center for Genomics Sciences, Universidad Nacional Autónoma de México, Av. Universidad s/n, Col. Chamilpa, 62210, Cuernavaca, Morelos, México.

**\*Corresponding author:** jfreyre@ccg.unam.mx (JAFG)

Genetic regulatory networks (GRNs) have been widely studied, yet there is a lack of understanding with regards to the final size and properties of these networks, mainly due to no network currently being complete. In this study, we analyzed the distribution of GRN structural properties across a large set of distinct prokaryotic organisms and found a set of constrained characteristics such as network density and number of regulators. Our results allowed us to estimate the number of interactions that complete networks would have, a valuable insight that could aid in the daunting task of network curation, prediction, and validation. Using state-of-the-art statistical approaches, we also provided new evidence to settle a previously stated controversy that raised the possibility of complete biological networks being random and therefore attributing the observed scale-free properties to an artifact emerging from the sampling process during network discovery. Furthermore, we identified a set of properties that enabled us to assess the consistency of the connectivity distribution for various GRNs against different alternative statistical distributions. Our results favor the hypothesis that highly connected nodes (hubs) are not a consequence of network incompleteness. Finally, an interaction coverage computed for the GRNs as a proxy for completeness revealed that high-throughput based reconstructions of GRNs could yield biased networks with a low average clustering coefficient, showing that classical targeted discovery of interactions is still needed.

## Introduction

Regulation is a critical biological process common to every living organism. Environmental cues such as nutrient availability and stimuli like temperature need to be sensed and integrated across a multi-layered decision-making system for an organism to mount an *ad hoc* response. A reductionist approach to biology has yielded extensive amounts of information about individual molecules and their interactions. However, it is now clear that most biological phenomena are complex and arise from the interaction of different components[1]. Transcriptional regulation is the process by which a set of regulator genes promote or inhibit the expression of other genes[2] The fact that regulator genes may influence other regulator and non-regulator genes allows for an convoluted network of interactions to be formed, thus enabling the integration of multiple signals by means of a differential flux of information through this network[3,4]. The information processing property of genetic regulatory networks (GRNs) demands a particular architecture governing the transcriptional circuitry, which must be shaped by evolution. A set of characteristics of GRNs such as a hierarchical-modular organization[5–9] and the existence of global regulators[10–12] have been proposed. Evolutionary studies so far have revealed that although organisms faced with similar environmental cues tend to show similar network motifs, most transcription factors have evolved independently, and do not share the same set of regulated genes[6,13]. It is important to note that selective pressure is known to act at the genome level, yet it is always through the consequences of functions and systems, and therefore can be studied through the commonalities present in different organisms regardless of whether they are genomic, proteomic or phenomic. Few studies so far have been able to assess conservation of GRN global properties across distant organisms[6] because they require analyzing the presence or absence of motifs, substructures, and properties in a range of phylogenetically distinct organisms whose reconstructed GRNs could be not available. Until recently, the small amount of reliable information on the transcriptional regulation of several organisms was a considerable challenge to infer evolutionary constraints on genetic regulation.

Currently, Abasy Atlas v2.0[14] (http://abasy.ccg.unam.mx) contains the most comprehensive collection of meta-curated bacterial GRNs having enough quality to allow system-level analyses. This repository also contains statistical and topological properties characterizing these GRNs. Additionally, Abasy Atlas classifies each gene as a global regulator, basal machinery gene, member of a functional module, or intermodular gene. This classification is based on predictions obtained by the natural decomposition approach[5,6,15]. To allow functional comparisons, modules are annotated by functional enrichment using gene ontologies as a controlled vocabulary. Meta-curation integrates various data sources and removes redundancy by 1) disambiguating gene symbols and 2) using a homogeneous representation for heteromeric regulators. Furthermore, GRN meta-curation allows unbiased comparisons among different versions of the GRN of an organism, and annotates each interaction as 'strong' or 'weak' according the experimental evidence supporting it[14]. This distinction provides information to define reliable gold-standards as GRNs whose set of interactions are only supported by 'strong' evidence[16]. Besides, Abasy Atlas v2.0 provides snapshots of the GRNs at different curation stages (historical reconstructions) for several organisms. A current limitation of Abasy Atlas is its biological diversity, it spans 42 bacteria (64% Gram positive and 36% Gram negative) distributed in only nine species (*Bacillus subtilis*, *Mycobacterium tuberculosis*, *Corynebacterium glutamicum*, *Escherichia coli*, *Staphylococcus aureus*, *Pseudomonas aeruginosa*, *Streptococcus pyogenes*, *Streptococcus pneumoniae*, and *Streptomyces coelicolor*).

In this study, we have used this database to detect a set of constrained properties that are likely shaped by evolutionary selection. Our results unveiled constraints in network complexity (density, number of interactions and number of regulator genes), providing strong statistical support to previous observations such as the presence of a long-tailed node degree distribution. The observed constraint in network density allowed us to estimate the total number of interactions that complete GRNs would have. Our framework represents the first approach based on biological information used to make this prediction and could be particularly valuable for integrative methods used to predict GRNs. The Dialogue for Reverse Engineering Assessments and Methods (DREAM) organized a challenge aimed at predicting GRN interactions based on expression data[17–19]. The participants were required to upload a list of possible regulatory interactions between *Escherichia coli* genes sorted by their confidence scores. This approach would benefit from the inclusion of prior information contained in the network structure. The ability to predict the number of interactions a GRN has, along with other topologically constrained properties, could be incorporated into these methods to allow for more accurate predictions. Using our predicted number of interactions, we computed an interaction coverage for each network, which we found serves not only as a completeness measure but also as a network quality indicator. The methodology developed in this study can be used to systematically assess whether other GRN topological properties have been subjected to evolutionary constraints, further enhancing our understanding of prokaryotic genetic circuitry.

**Results**

***Constraints in the properties of the Abasy Atlas GRNs***

GNRs possess a series of topological properties that allow for the correct integration of, and response to, environmental signals. The density of a network is a property quantifying the fraction of existing interactions relative to the total number of possible interactions given the number of genes (see Materials and methods). If GRNs share some common structural patterns, we could expect to see a convergence or trend of their densities into a defined range. Interestingly, we found that Abasy GRNs (N = 71) densities followed a trend towards a relatively small value as the number of genes in the GRN increased. Networks with less than 500 genes showed a significant amount of variation but followed a nonlinear trend towards a value of low density (**Fig. 1a**). We found that this nonlinear trend follows a power law ($d \sim n^{-\gamma}$) with $\gamma$ = 0.78 ≈ 1 ($R^2$ = 0.96), strongly suggesting that a hyperbolic (inverse) behavior governs the relationship between density and number of genes. The same behavior was observed using a set of non-redundant Abasy GRNs ($\gamma$ = 0.76, $R^2$ = 0.95) (**Supplementary Fig. 1**).

Differences in the curation methods or reconstruction strategies could account for this result, but this trend was conserved when analyzing a set of historical reconstructions (see Materials and methods) of *E. coli*[20,21] and *Corynebacterium glutamicum*[15] GRNs. Furthermore, four different reconstructions of the *Bacillus subtilis* GRN each constructed independently with different curation strategies[22,23] including a computational prediction[24] continued to show a trend towards a low density (**Supplementary Fig. 1**). This trend was observed when GRN genomic coverage (a traditional measure for completeness; the fraction of the organism genome that is included in the GRN reconstruction), or a set of only non-redundant networks (see Materials and methods) were used (**Supplementary Fig. 1**). Indeed, changes in GRN completeness seemed to negatively correlate with density (**Fig. 1b note the density axis scale**), suggesting that density is likely constrained in the complete GRNs.

The variability observed in highly incomplete GRNs (genomic coverage < 0.25) can be explained by the fact that a network with such a low number of genes could not possibly have a density as low as observed for more complete networks without having disconnected genes, which are not included in GRN reconstructions (e.g., a network with 200 nodes and a density of 0.001 would have ~40 interactions which, in the best case scenario, would leave 120 disconnected nodes). To illustrate this, we generated random subnetworks (using a snowball sampling approach, see Materials and methods) of the most recent *E. coli* GRN (511145_v2017_sRDB16_dsRNA) and assessed the sampled networks densities excluding disconnected nodes. As expected, the relationship between subnetwork mean density and number of nodes followed the same trend as observed for the networks in Abasy. Accordingly, the variation increased as subnetwork size decreased (**Supplementary Fig. 1**). In fact, normalizing network density by its expected value derived from *E. coli* random subnetworks unveils the putative invariance in density (**Supplementary Fig. 1**).

GRNs are usually modeled as directed graphs because only a set of regulator genes may activate or inhibit other genes. Therefore the patterns observed in density could be understood if the number of regulators, and hence the number of interactions, in the networks are being constrained. Notably, the number of regulators in the studied GRNs showed a high correlation with the number of genes or genomic-coverage (**Fig. 1c and Supplementary Fig. 2**). This trend was also observed in historical (*E. coli*, but not *C. glutamicum*) and independent GRN reconstructions (*B. subtilis*) (**Supplementary Fig. 2**). These observations suggest that the evolutionary constraints that have shaped the structure of prokaryotic GRNs constrains the percentage of genes that can act as regulators (~7% in average, **Fig. 1c**). Although the discovered patterns in network density and number of regulators are very likely to be related, it is still elusive whether one and which, is a causal agent for the other or whether unknown topological constraints confound both.

### *The complexity of GRNs could be bound by their stability*

We found that the relationship between density ($d$) and number of genes ($n$) of Abasy Atlas GRNs follows a power law ($d \sim n^{-\gamma}$) with $\gamma \approx 1$ (**Fig. 1a**). Interestingly, this result suggests that GRN complexity could be bound by the number of genes as predicted by the May-Wigner stability theorem[25]. Robert M. May concluded that the stability of randomly connected systems depends on the number of variables (*n*), connectance (*C*), and interaction strength dispersion ($\alpha^2$)[25]. The May-Wigner stability theorem says that randomly connected large systems are stable if $nC < 1/\alpha^2$ (see **Supplementary note 1** for some additional details on this theorem and a brief historical summary). Connectance is analogous to density in graph theory as both quantify the fraction of existing interactions relative to the total possible (hereafter we use connectance and density as synonyms, but we prefer connectance when referring to early works or ecological communities where this term is standard).

Bacterial GRNs exhibit a hierarchical modular organization as predicted by the natural decomposition approach[5,6,8,15] and other studies[7,8,26]. As system stability is a requirement for organism survival, there must be constraints shaping the modularity and number of hierarchical layers of GRNs. Besides the inverse relationship between density and number of genes (*nd* = *k* = 0.40), we also observed that the interaction strength dispersion of Abasy Atlas GRNs is a constant ($\alpha^2$ = 1/*k* = 2.50) (**Fig. 1a**). The same result was also observed when we used a set of non-redundant Abasy GRNs (*nd* = *k* = 0.37, $\alpha^2$ = 1/*k* = 2.70) (**Supplementary Fig. 1**). This shows that GRNs stability is ensured if the interaction strength dispersion is at least 2.50 (*nd* < 0.40). Conversely, empirical data from food webs also have shown a hyperbolic behavior governing connectance and number of species but the interaction strength dispersion falls into a range with lower values (1/6 < $\alpha^2$ < 1/2)[27–29].

Further research is needed to understand how evolution is acting upon the GRNs organization constraining modularity and hierarchy to ensure stability, and to evaluate how the probability of stability constrains the landscape of possible GRNs structures (the GRNs organizational landscape[15]).

## *An analysis of GRN properties shows that the organization of complete prokaryotic GRNs is not random but scale-free*

The trend towards a relatively small density value implies that prokaryotic GRNs are sparse. This observation is likely a result of a set of common organizing principles. The existence of highly connected global regulators in a GRN with such a low density should cause non-global regulatory genes to have a low number of interactions (relative to global regulators). Thus, the average node connectivity should be low. In fact, a previous study found that in sparse complex networks (a social network in this case) the node degree distribution changes as link density increases; these sparse complex networks initially showed a power-law node degree distribution, but as link density increased a divergence from power law was noted[30].

The power-law behavior governing the node degree distribution has been proposed as a common organizing principle of GRNs[6]. For the sake of simplicity, we focus here on a general P(k) distribution combining both P($k_{in}$) and P($k_{out}$) distributions. In the literature of network biology, there has been a debate when referring to 'universal' topological properties of biological networks[1,31], especially on whether a specific probability distribution governs GRN P(k). It has been particularly stated[32] that sampled Erdos-Renyi (ER) networks (with an initial Poisson governed node degree distribution) could present *power-law* P(k) distributions. Thus, creating the misleading idea of sampled random networks being hierarchical structures with global regulators (hubs). Furthermore, these results raised the possibility of biological network P(k) long-tailed distributions to be an artifact of sampling (i.e., incompleteness).

The main problem with the aforementioned (and other) studies is the assessment of the goodness of fit to a long-tailed distribution by only using the coefficient of determination for a linear regresion[31]. To robustly find whether the analyzed GRNs P(k) follow a long-tailed distribution, or if previous observations[1,9,15] could be in fact an artifact of sampling[31,32], the goodness of fit of GRN connectivity to different probability distributions was assessed through computing the Kolmogorov-Smirnov (KS) D statistic (distance) using a maximum-likelihood-estimate (MLE) fitted theoretical distribution, and by calculating the log-likelihood ratio test of the MLE fitted distributions against a power-law distribution[33].

The log-likelihood ratio tests showed a preference for a power-law distribution over exponential and Poisson distributions for the GRNs studied, and rendered no significant distinction between lognormal, stretched exponential and power law. A slight preference for truncated power law was also evident (**Fig. 2a, b**). Furthermore, the KS D statistic favored the power law and other long-tailed distributions as they had the smallest difference between the data and the model (**Fig. 2a**). No GRN P(k) showed a good fit to a Poisson distribution, while a set of sampled[32] (see Materials and methods) ER graphs parametrized to have an average probability and size equal to the GRNs, had the best P(k) fit to a Poisson distribution as demonstrated by the calculated KS D statistic (**Fig. 2a**)[33,34].

Previous reports[32] claimed that the smaller the subsample of the ER networks analyzed, the better the fit to long-tailed distributions. It is important to highlight that we initially observed a preference for a power law (over a Poisson) for **all** ER-graphs' P(k) distributions (**Supplementary Fig. 3**). While unexpected, this observation can be explained by a) the broken assumption when comparing Poisson and power-law distributions due to the fact that the likelihood ratio test assumes nested distribution (Poisson is not nested within power law unlike the other distributions tested), or b) the effect on the data of a fitted parameter (*xmin*) needed for the MLE of a power-law distribution[33,34]. Because we observed similar results for the KS D statistic (**Supplementary Fig. 3**), we assumed the latter to be the cause of the observation, as the KS_D statistic does not assume nested distributions. The *xmin* parameter finds the minimal accepted value of connectivity for which the power law distribution is valid and causes ER graphs derived P(k) to appear power-law like (see **Supplementary note 2**). When this analysis was performed fixing the value of *xmin* to a value of one (therefore forcing the fit to consider all the data) the ER P(k) preference for power-law distributions (specially for the most complete ER networks) disappeared (**Fig. 2a, b**).

While we were able to replicate previous results showing a better fit to a power-law distribution with more incomplete ER graphs[32] (**Fig. 2c**), when comparing the fit of these ER networks to Poisson and power law, an evident and significantly better fit for a Poisson P(k) distribution was observed even for the most incomplete networks (**Fig. 2c-e** and **Supplementary Fig. 3**). Furthermore, the smallest available GRNs showed a strong deviation from a Poisson distribution, and, as previously reported[30], we detected a negative correlation between power-law fit and network density (**Supplementary Fig. 3**). These observations suggest that a Poisson distribution is not a *bona fide* model for deriving GRNs P(k) (**Fig. 2d and Supplementary Fig. 3 and 4**). Long-tail distributions explain better the prokaryotic GRNs P(k) here studied, thus supporting the existence of hubs and a non-random organization as properties, rather than as artifacts of sampling[5–9].

The previously discussed[32] effect of sampling on ER networks (**Fig. 2d**) motivated us to assess the possibility of further rejecting ER networks as a model for GRNs. If GRNs are not derived from ER graphs, then some of the properties observed in biological networks should be incompatible with the ER model. Mainly, biological networks have been proposed, and seem to have, long-tailed P(k) distributions[1,8]. While some instances of a Poisson derived network may show a high goodness of fit to a long-tailed P(k) distribution (**Fig. 2**), we hypothesize a measure of P(k) *tail length* to show higher values in biological networks as opposed to comparable ER graphs. Therefore, we defined tail length as the common logarithm of the difference between the maximal and minimal degree of a network.

A GRN comparable (see Materials and methods) set of ER graphs was constructed (**Fig. 3 a, b**), and we studied the tail length and average clustering coefficient distributions of this null subset compared to Abasy GRNs. We also applied this methodology to Barabasi-Albert (BA) growing random networks model. Both average clustering coefficient and tail length distributions from GRNs were different from their parametrized ER and BA counterparts (**Fig.3 c, d**). The construction algorithm for the BA graphs caused a significant difference between the null model's density and the GRNs density (**Fig. 3b**). Thus, although the average clustering coefficient differences are evident, no conclusion can be drawn regarding them (see Discussion). The fact that BA networks are not easily comparable with the GRNs is interesting as BA networks may intuitively be considered as a better model for GRNs because of its power-law properties. The significant difference between average clustering coefficient and P(k) *tail-length* between GRNs and their equivalent ER graphs (**Fig. 3 c,d**) suggest that Abasy GRNs cannot be derived from an ER model, given their densities, average clustering coefficients and tail lengths (**Fig. 3**).

Furthermore, when analyzing the properties of ER networks sampled to appear *power law*, their tail-lengths and clustering coefficients were further incompatible with those of the biological networks (**Fig. 3**; compare e,f with c,d). These combined results provide further evidence suggesting that a Poisson distribution is not likely to model the distribution of complete prokaryotic GRN P(k) appropriately, and that the observed properties (e.g., the existence of hubs causing long tailed P(k) distributions) are not likely an artifact of GRN incompleteness. Importantly, this methodology can be easily extended and modified to assess the plausibility of other network null models for the existent GRNs possibly aiding in the design or parametrization of algorithms to infer GRNs.

*Estimating the number of interactions of complete GRNs*

A discovered completeness-density trend (**Fig. 1a**) should imply a relationship between the number of genes and the number of interactions a network has. In fact, this was observed amongst the networks in Abasy (**Fig. 4a**) and when analyzing GRNs normalized by genomic coverage, historical reconstructions and a subset of non-redundant networks (**Supplementary Fig. 5**). Both the completeness-density and completeness-number of interactions correlations (**Fig. 1a and 4a**) enabled us to generate predictive models for inferring the number of interactions a GRN would have (see Materials and methods). Briefly, by incorporating either the relationship between number of genes and edges (edge regress model (EdR), **Fig. 4a**), by assuming an invariant density (density invariance model (DI), **Fig. 4b and Supplementary Fig. 1**) or incorporating the trend in density (density proportionality model (DP), **Fig. 4c and Supplementary Fig. 5**), we could estimate a set of proportionality factors that explained the number of edges (interactions) in terms of the number of nodes (genes).

Assuming the total number of nodes of a GRN to be the number of annotated genes in the genome of that organism, we could generate three models (**Fig. 4a-c and Supplementary Fig. 5**) to estimate the total number of interactions. A comparison between some organisms GRN number of genes, genomic coverage, actual number of interactions and total estimated number of interactions predicted by each model is given in **Table 1** (see **Supplementary Table S1** for a full listing of all organisms in Abasy).

Of the three models, both the EdR and DP approaches had the highest goodness of fit, as they incorporated variance from the smallest GRNs. Both have only a free parameter to estimate, and an error parameter, but EdR assumes a simpler underlying model. Interestingly both model predictions are in agreement with each other, while the DI model predicts more interactions, the difference could be accounted on the fact that it was built using a reduced number of GRNs (non-redundant and highly complete networks).

To assess the capability of these methods to model the number of interactions of the complete networks, we decided to test their ability to predict the progressive increase on the number of interactions of the historical reconstructions of *E.coli* which have a relatively high number of genes (or genomic coverage) when compared to all Abasy GRNs. All the models had a good fit to the data ($0.87 \leq R^2 \leq 0.91$), with small differences between them. Notably, the DI model had the best fit, despite having the poorest fit to the most incomplete networks in Abasy (**Fig. 4b**), the DP and EdR models were very similar (**Supplementary Fig. 5**). These results are expected as all the historical reconstruction GRNs had a relatively high genomic coverage, and the DI model in fact considers information from only the most complete networks without incorporating variance from the smallest GRNs. Until today, few efforts have been made to estimate GRN size considering missing interactions[36], and our models are the first ones to integrate meaningful biological data consistent across distinct organisms. We leveraged our capability of finding putative trends or constraints on the topology of GRNs to make further predictions about their final topology. The trends in the number of regulators, network density and the total number of interactions could be valuable for the still ongoing development of methodologies aimed at predicting complete GRNs *de novo* by integrating high-throughput data.

### *Assessing GRN completeness based on number of interactions*

We next assessed GRN completeness using an interaction coverage, rather than genomic coverage score. Notably, the interaction coverage computed from our models showed a high correlation with genomic coverage ($R^2$ = ~0.88) and with average clustering coefficient ($R^2$ = ~0.60, **Supplementary Fig. 6**), thus suggesting that more complete networks tend to have a higher average clustering coefficient.

To assess differences between genomic and interaction coverages, we computed a *comparison score* defined as the natural logarithm of the ratio between interaction coverage (calculated independently with each prediction model) and genomic coverage. Negative values of this score indicate that the interaction coverage is penalizing (i.e., predicting the GRN to be less complete) when compared to genomic coverage. We expected networks with a higher average clustering coefficient to be less penalized by the interaction coverage, as they contained more interactions ($R^2$ = 0.40, **Supplementary Fig. 6**). A dependency between the computed *comparison score* and average clustering coefficient was evident in two predictive models (EdR and DP **Fig. 4d, f**) as the less penalized networks were the ones with a higher average clustering coefficient. Notably, the model based on a density invariance (DI) fails to recapitulate this result. (**Fig. 4b,e**). In all cases, networks were mostly penalized as evidenced by negative penalization values (**Fig. 4 d-f**). Thus far, the analyses were carried out using all networks available in Abasy including different organism historical, independent, and meta-curated GRNs. The density invariant assuming model developed herein uses only a set of non-redundant networks with a number of genes higher than 1000 to identify the putative invariant density (a smaller training set).

We also evaluated whether network redundancy could be biasing our estimates for the other two estimators. We re-estimated model parameters using only a set of non-redundant GRNs (i.e. using the most complete GRN per organism). The parameters and results were similar regardless of the analyzed set (**Supplementary Fig. 7 and Supplementary Table S2**). The overall negative penalization scores suggest that most networks still lack a significant amount of curation for discovering intra-modular interactions. Furthermore, our computed interaction coverage can be used to assess network completeness and quality, as deviations from this pattern may reveal reconstruction biases (see below).

## Curation of high throughput experiments could bias GRN discovery as revealed by subsetting analysis of a GRN gold standard

We found that the interaction coverage and penalization score estimates for three networks of *Mycobacterium tuberculosis* (83332_v2015_s15, 83332_v2016_s11-12-15 and 83332_v2018_s11-12-15-16) were strong outliers in our previous analyses, particularly in the dependency between average clustering coefficient and completeness or *comparison scores* (**Fig. 4d,f and Supplementary Fig. 6,7**). These networks are based mainly on the 2015 reported network reconstructed by high-throughput experiments, which presents a very low average clustering coefficient (~0.12) indicating a low network modularity (**Supplementary Fig. 7**)[37,38]. Furthermore, one of these networks showed an interaction coverage above one, but a low average clustering coefficient (**Supplementary Fig. 6**). To address whether a biased reconstruction based mainly on high-throughput methods could be the cause of our observations, we created three subsets of the *E. coli* gold-standard GRNs[21] consisting of interactions supported only by high-throughput curated experiments (HT), interactions with evidence from non-high-throughput experiments (non-HT), and a last one containing both. Although all of them contained the same set of nodes (and hence the same genomic coverage), the non-HT subnetwork contained a higher number of edges than the HT. This difference in the number of regulatory interactions could be due to the great amount of curation this regulatory network has. However, no differences among the intrinsic properties of these subnetworks should be observed. We hypothesized that the HT reconstruction contained fewer modular interactions if experiments were performed on a set of regulators yielding a poorly interconnected tree-like structure. Effectively, the average clustering coefficient of the HT subgraph was ~5-fold less than the non-HT subgraph and ~7.7-fold less than the combined (i.e., containing both HT and non-HT interactions) subgraph (**Fig. 5**).

To analyze if this reduction in the average clustering coefficient is due to the structure of the graph rather than a diminished number of edges, we performed a random removal of edges of the combined subgraph until it contained the same quantity as the HT; this was repeated 1000 times to have a random distribution of clustering coefficient averages. Interestingly, the distribution of the average clustering coefficient of the randomized networks was significantly higher from that of the HT subgraph (Z = -3.9, p < 0.001)(**Fig. 5**), and significantly lower than the experimental curation and combined subgraphs (Z = 9.3, p < 0.001, and Z = 18.4, p < 0.001, respectively). Thus, implying that the high-throughput GRN structure has arisen from a specific type of sampling causing a particular organization, specifically yielding a low average clustering coefficient while maintaining the genomic and interaction coverage in the HT reconstruction. This phenomenon holds true for curated regulatory networks, as revealed by analyzing the average clustering coefficients of the GRNs available in Abasy Atlas[14].

Interestingly, the most complete networks were the ones based mostly on manual curation (e.g., *E. coli GRNs*).[21–23,39] These networks showed a higher average clustering to coverage ratio than the ones based mostly on high-throughput[40], computational predictions[24] or meta-curations[14] (**Supplementary Fig. 7**). When analyzing a non-redundant subset of Abasy GRNs, those with a higher genomic coverage were usually the ones with higher average clustering coefficient except for 83332_v2015_s15 (the previously discussed *M. tuberculosis* HT GRN, **Supplementary Fig. 7**). Overall, these observations suggest that the genomic coverage is not the best proxy for network completeness provided an interaction coverage to be available, and that average clustering coefficient could serve as a network quality indicator, as very low values of average clustering coefficient could indicate a biased GRN reconstruction. Furthermore, with the upcoming of high-throughput and computational discovery of GRNs, an increase in the genomic and interaction coverages of networks is expected, potentially creating a misleading belief of network completeness. Reaching high levels of genomic coverage does not necessarily represent a highly complete network, and an integrative estimate of the number of missing interactions and clustering coefficients is needed not only for assessing completeness but also for guiding the experimental or inferential strategies to complete GRNs.

**Discussion**

The present study was powered by the availability of many meta-curated GRNs present in a single database[14]. By simultaneously comparing properties between these GRNs, an implicit evolutionary study was conducted revealing a set of constrained properties. The inherent assumption underlying our conclusions is that a trait (or in this case a network property) present in phylogenetically distinct organisms would have most likely appeared in a common ancestor and prevailed through natural selection[6] or other evolutionary forces. Recent studies support this assumption, showing that the functional architectures of disparate bacteria are conserved by convergent evolution thus suggesting that bacterial GRNs evolved in a constrained organizational landscape[6,14,15].

Because of the lack of available complete GRNs, our results compared organisms with a heterogeneous amount of information. This represents a potential source of noise or bias which would complicate the elucidation of constrained properties. To account for the variation in existing regulatory information, the properties or traits were always studied in relation to network completeness using the number of genes or network genomic coverage as a quantitative proxy for it. Using the approach described here, the systematic analysis of constrained properties in prokaryotic GRNs is feasible.

Abasy Atlas v2.0, the database used in this study to obtain regulatory information and properties, contains 71 distinct regulatory networks covering 42 strains of nine different species. The implicit redundancy of the GRNs present in this database could, through *pseudo replication*, bias our results. This was accounted for by repeating all analyses based on comparing between GRNs using a non-redundant subset of the networks present in Abasy. Furthermore, the usage of historical reconstructions[20,21] and independent reconstructions[22–24] of regulatory networks to validate our results is a strategy that allowed us to further assess the existence of observed properties, and to exclude curation strategies and network completeness as sources of artificial observations. It is not yet clear to what extent and why, the observed properties are being constrained by evolutionary forces. We hypothesize that the observed constraints, including network complexity and number of regulators, could be explained by evolution selecting for system stability and the existence of regulatory motifs which enable cells to perform computations to integrate differential signals[4], preventing GRNs from randomly growing or losing a defined structure[6,8,15]. Future studies could use similar approaches as the ones described herein to test this hypothesis.

Network biology has revolutionized life-science oriented research. An increased understanding of any organism could be gained through the analysis of the different systems that compose it. Nonetheless, topological analyses of biological networks have faced strong controversies arising from hyperbolic claims stating universal properties of all biological networks[1,31]. In the present study, we have shed light on the polemic subject of GRN degree distribution[32]. Through rigorous statistical approaches we have shown that long-tailed distributions (which favor the existence of hubs and a non-random organization), and not Poisson, better explain the degree distributions in bacterial GRNs. Furthermore, we demonstrated that ER networks that appear *power-law* due to an artifact of sampling still have a better fit for a P(k) Poisson distribution, and their average clustering coefficients and tail-lengths are highly incompatible with currently known GRNs. Notably, remarks referring to the extremely low average clustering coefficient of these sampled ER networks have been published[41]. Here we extend these observations by actually comparing *power-law like* ER network properties with meaningful GRNs.

Although a limitation of one of our approaches was the invalidity of comparing power-law and Poisson distribution fits using likelihood ratio tests, our observations are further supported by observations of the KS D statistic, which is free of any nested model assumption. Moreover, an approach based on constructing ER and BA networks parametrized to follow our GRNs densities allowed us to reject the possibility of our biological networks as being derived from these models based on their paired densities, clustering coefficients and tail length distributions. We consider the incapacity to construct BA networks parametrized to fit our biological network properties evidence of how unlikely it would be for our GRNs to be derived from them. This same approach can be used and extended to assess the consistency of GRN inference methods by systematically attempting to find incompatible properties between the predicted GRNs and the distribution of their curated counterparts.

The discovered constraint on GRN complexity was exploited to find a relationship between number of genes in a GRN, and number of edges, and as a consequence predict total number of genetic interactions. These predictions use the trends discovered herein to produce an expected number of interactions, along with 95% confidence intervals. The current challenges existing in curating or predicting GRNs from high-throughput data make the estimate of the total number of expected interactions for the complete networks a key factor allowing algorithms to set boundaries when assessing the possibility of connections between all possible gene pairs[17,19]. A prediction based on observed properties of the amount of interactions a complete GRN, has not been, to the best of our knowledge, described before. It represents an important contribution to the field as it allows databases and curators to gain an idea of how truly incomplete current GRNs are[36]. For example, estimates on the amount of missing information suggest that half of the *E. coli* genetic regulation is still unknown[42]. We extend these predictions and present quantitative approaches for estimating the amount of total (and thus also of unknown) interactions. We computed an interaction coverage score based on the predictions made by our different models and discovered that most GRNs are more incomplete than previously thought based on genomic coverage.

Interestingly, networks with a higher average clustering coefficient were penalized the least (when compared to genomic coverage) by two of our three models. The outliers to this observation were networks based on a high-throughput reconstruction of *M. tuberculosis* GRN[40]. We provided a proof of principle, by subsampling a gold-standard GRN[21,43], of how high-throughput curation could bias GRNs reconstruction making them appear complete in terms of both genomic and interaction coverage, but presenting a low average clustering coefficient.. Our results suggest that high-throughput curation of GRNs could yield networks with modularity lower than expected. This result aimed at raising the possibility of biases existing when curating GRNs mostly from high-throughput experiments, we acknowledge that a more detailed revision of GRNs and high-throughput technologies, and their biases, should be performed. Overall, our observations suggest and enable the finding of global structural properties constrained in GRNs, which can be used to understand how evolution has shaped their topology, aid in predicting other properties (i.e., number of interactions a complete GRN would have) and will be particularly valuable for guiding GRN inference and prediction algorithms.

**Conclusion**

In this study, the availability of a large collection of meta-curated GRNs in a single database enabled the analysis of different properties unveiling topological constraints, which we hypothesize to have underlying evolutionary causes. We have found GRN density, number of regulators and number of interactions to have a constrained space of possible values (organizational landscape); in the latter two cases with a strong relationship with the number of genes in the network. Our results suggest that bacterial GRNs node degree distributions are governed by long-tailed distributions, supporting the existence of global regulators and a non-random organization. We could discard ER and potentially BA as faithful representations of GRNs given their densities, average clustering coefficients, and tail-length values, settling the current debate of whether degree-distribution claimed properties exist or are an artifact of network sampling.

Three different estimations of GRN total number of interactions were computed, which represent a valuable tool that can be used to aid in network inference. Most GRNs were penalized when comparing their interaction coverage with their genomic coverage. Interestingly, the least penalized networks were those with a high average clustering coefficient, except for a set of networks all based on a high-throughput reconstruction of the *M. tuberculosis* GRN. Finally, our results suggested that high-throughput based curation could bias GRN discovery yielding tree-like networks with a low average clustering coefficient. Nonetheless, a more thorough analysis of high-throughput methodologies and their resulting reconstructed GRNs is needed to detect whether this pattern holds true for all high-throughput based GRNs. The methodology presented here can be used to systematically find constrained topological properties throughout the GRNs available in public databases, and aid in the developing area of GRN *de novo* inference or prediction.

**Materials and methods**

*Data retrieval processing and availability*

Prokaryote GRNs were obtained from Abasy Atlas v2.0, a database that contains 71 reconstructed and meta-curated GRNs covering 42 bacteria. All the analyses were performed on all 71 GRNs unless stated otherwise. All analyzed data is available as downloadable files from Abasy Atlas at http://abasy.ccg.unam.mx. Transcriptional regulation was modelled as a graph, where nodes represented genes, and the edges between them represented the existence of a regulatory interaction. For the analyses of number of regulators, directed graphs were used and regulators were defined as the set of nodes with an out-connectivity > 0 (i.e., they are a known regulator of at least one gene).

*Density analysis*

For a given network, density is defined as the number of actual edges of the network over the number of potential total edges. In GRNs which are represented as directed graphs it is calculated using the formula:

$$I_{MAX} = |N_G|^2 - |N_G| \tag{1}$$

$$D_G = \frac{|E_G|}{I_{MAX}} \tag{2}$$

Where $D_G$ stands for the density of graph G, $E_G$ and $N_G$ are the sets of interactions and nodes in graph G, respectively, and $|x|$ is the cardinality of set $x$. This formula was used to estimate the densities for all GRNs present in Abasy Atlas.

*Computing average clustering coefficient*

The clustering coefficient is a node-level measurement of modularity. An easy analogy to understand clustering coefficient is a social network, where nodes represent people and edges represent the existence of a friendship between two nodes (people). The clustering coefficient quantifies for a given person how many of its friends know each other. A clustering coefficient of one would indicate a high centrality, within a module (group of friends), of that node (person) meaning that all of this person's friends are also friends with each other. Formally, the clustering coefficient is calculated as (assuming an undirected graph):

$$C_v = \frac{N_v}{\frac{I_{MAX_v}}{2}} = \frac{2N_v}{I_{MAX_v}} \qquad (3)$$

Where the numerator ($N_v$) represents the number of actual edges (interactions) between the current node (*v*) neighbors, and the denominator is the maximum number of interactions the neighbors could have. In this study, the clustering coefficients of all nodes in a given graph were averaged (arithmetic mean).

*Historical, independent and non-redundant GRN reconstructions*

We leveraged the availability (within Abasy) of different networks curated or constructed for the same organisms. In this study, we defined the set of different public versions of a network (e.g., the different versions of RegulonDB for *E. coli*) as historical reconstructions. In contrast, networks from the same organism reported from different databases or sources (e.g., Subtiwiki, DBTBS and the *in silico* reconstruction by Arrieta *et al.* (2015)[24] for *B. subtilis*) were considered independent GRN reconstructions. Finally, given the existence of several non-independent GRNs for a given organism we defined a set of non-redundant GRNs to ensure the results reported herein are not an artifact of *pseudoreplication*. The non-redundant GRN set was constructed by selecting the most-complete GRN per organism available in Abasy.

*MLE and KS D estimation*

To estimate the parameters for the compared P(k) distributions, a vector of degrees was obtained for each GRN. This vector was used as the input data to fit the different probability distributions using maximum likelihood estimates found with in-house scripts and the library powerlaw for python [33] based on methods previously described [34,44]. A loglikelihood ratio test was used to compare the goodness of fit of the different distributions versus a power-law distribution. The scores were plotted on a heat map using matplotlib and seaborn. As another measure of goodness of fit, the Kolmogorov-Smirnov D statistic was also calculated and depicted as a heat map. The MLE methods used require an '*xmin*' parameter for the power-law distribution. This parameter selects the minimal degree value from which a power-law would have its best fit. Because this parameter induces trimming of the data causing non-long-tailed data to have a good fit to a power law distribution, these analyses were repeated with a fixed *xmin* value of one, thus including all of the data (see **Supplementary note 2**).

*Incompatibility between null models and biological GRNs*

Randomized ER or BA graphs were generated to follow the observed GRNs density and size distributions. These 1000 random networks were considered to be analogous to biological GRNs, and their average clustering coefficients and[45] tail lengths (**defined in main text**) were calculated and compared with the biological networks. If the models were a faithful representation of the GRNs, then no significant difference between any properties should be observed, if the contrary occurred then the model was considered incompatible given the networks densities and clustering coefficients, or given the networks densities and tail lengths. Notably, this methodology can be modified to accommodate other null models with construction parameters different than density.

*GRN comparable Erdos-Renyi* **graphs**

An ER graph is a network that can be described by a characteristic node whose connectivity is the expected value of the whole graph's P(k) following a Poisson distribution. The parameters needed to construct them are the number of genes and the average connectivity. GRN density was used as a proxy of network average connectivity, a random sampling from the GRNs densities and sizes was used to create 1000 ER graphs. Average clustering coefficient and tail-length distribution means were compared with a Mann-Whitney U test, and statistical significance was considered at a significance level $α < 0.05$.

*Graph* **sampling methods**: *Information retrieval sampling*

While originally designed to model the discovery of protein-protein interaction networks, we decided to use this sampling framework as it was previously reported to show Erdos-Renyi graphs' P(k) to present a good fit to a power-law distribution. The approach is extensively described in Han *et al.* (2005). Briefly, two parameters, bait and edge coverage are used for sub-setting a network. In our scenario, bait coverage represents the fraction of genes whose regulatory interactions will be included and edge coverage represents the fraction of existing interactions per bait (simulating technical and experimental limitations) whose paired genes will be also included in the GRN. This approach was used to sample the graphs on the results unless otherwise stated.

*Graph sampling methods: Snowball sampling*

This sampling algorithm is very likely to accurately model curation of GRNs. It is based on the idea of reconstructing a network through layers of connected components by following a breadth-first search (i.e., by discovering all of the neighbors and the neighbors of the neighbors … of a seed node). Briefly, a random gene is selected to serve as a seed. The gene neighbors (interactors) are added to the network until the final size is achieved. If all the seed neighbors have been included, then the next layer of neighbors, following a breadth-first search, is systematically added until the desired percentage of completeness is achieved. If the network has several disconnected components and the desired sample size has not been achieved, another gene is chosen randomly as seed to continue with the sampling procedure. This approach could mimic the classical curation of GRNs as information is retrieved from experiments that tend to be performed based on other known interactions.

*Number of interactions and Interaction coverage estimation*

Two different, yet related network properties (density and number of interactions) seemed to show a trend or dependency with the number of genes in the network. Although there is a direct mathematical formula linking density, number of nodes and number of edges, including both relationships (density-number of nodes, and number of nodes-number of edges) under the same model is not trivial, primarily because of the quadratic dependency between the number of nodes, number of edges and graph density. We therefore devised three independent approaches, each incorporating our observations to predict GRN total number of edges.

*Density proportionality approach*

We assume a conservation of the trend observed in **Fig. 1a** between network *completeness* and density and modeled it using an exponential decay fit (**Supplementary Fig. 4e**). Thus, the tendency GRNs densities follow should be governing the number of interconnections these networks have. Below follows a derivation of how assuming a specific value for network density can allow for a total number of edges prediction. First, let us rewrite equation 2 to define the number of interactions a network has in terms of its density and maximum possible interactions:

$$I_{MAX} \cdot D_G = |E_G| \qquad (4)$$

Where $D_G$ represents the density for graph G, $E_G$ the set of interactions in graph G and $N_G$ the set of nodes (genes) present in graph G. $I_{MAX}$ represents the maximum number of edges a genetic regulatory network would have given its number of nodes. This equation represents a relationship between the number of interactions and the number of nodes given a specific graph density. Using ordinary least squares regression (OLS), it is possible to estimate the proportionality density factor $D_G$ given the networks in Abasy (**Fig. 4b**) by modelling the change in density using a linearized exponential decay function (**Supplementary Fig. 4e formula**). The proportionality factor is therefore defined as

$$D_G = e^{-\alpha \log(N_G) + \varepsilon} \tag{5}$$

Where $N_G$ corresponds to the number of genes in the genome and $\alpha$ and $\varepsilon$ represent the coefficient estimates for the linearized exponential decay model. This proportionality factor can be combined with a predicted number of genes in a GRN to predict the number of interactions it would have. We decided to use genome annotation information based on ORFs as a proxy for the number of genes each GRN in Abasy would have if complete. Knowing the total number of genes a complete GRN would have and the determined density proportionality factor, we effectively computed a total number of interactions prediction for each GRN

$$|I_G| = I_{MAX} * D_G \tag{6}$$

$$|I_G| = I_{MAX} * e^{-\alpha \log(N_G) + \varepsilon} \tag{7}$$

*Density invariance approach*

For the density invariance (DI) approach eq 6 is used, but $D_G$ is the mean density of the set of non-redundant networks in Abasy with at least 1000 genes. This was decided because of the biases that very incomplete networks have showing a higher density than most complete networks (**Fig. 1a** and **Supplementary Fig. 1**). This model assumes that the GRN density is constrained to a very small value (obtained as the mean density of the most complete GRNs) therefore bounding the number of interactions the complete GRNs would have.

*Edge regression approach*

The linear and robust correlation between the number of genes and the number of interactions in our analyzed prokaryotic GRNs motivated this approach. The correlation from a non-redundant set of Abasy GRNs was modeled as a linear dependency between the two variables, and the coefficients were estimated by OLS (**Fig. 4a**). Genome annotation information based on open reading frames (ORFs) was obtained from Abasy for each of the analyzed networks and used as a proxy for the number of genes in the genome. Finally, the extrapolation of the observed dependency was used to generate the number of interactions prediction for each GRN.

$$|I_G| = \alpha N_G + \varepsilon \tag{8}$$

In all three approaches interaction coverage was defined as the number of edges present in the GRN over the predicted total number of interactions in the GRN

*Comparison between interaction and genomic coverage*

A score that enabled direct comparison between our interaction-based completeness estimates and the classical genomic coverage estimates was implemented as:

$$C_g = \ln \frac{I_{cov}(g)}{G_{cov}(g)} \tag{9}$$

where $C_g$ is the comparison score for GRN $g$, $I_{cov}(g)$ and $G_{cov}(g)$ represent the interaction and genomic coverages (respectively) of $g$. If the same completeness is estimated from the interaction and genomic coverages, then this score will have values close to zero. Negative values indicate that the interaction coverage is predicting a GRN to be less complete in comparison to the estimate of genomic coverage and vice-versa. This score was calculated for all GRNs present in Abasy for the three different interaction coverage scores (one per predictive model, see above).

*High throughput E. coli analysis*

*E. coli* GRN metadata was processed to obtain the different experimental evidences for each of the interactions present. Three subsets of this GRN were created: 1) Containing both high throughput and classical experimental supported interactions (All), 2) Only classical experimental supported interactions (non-HT) and 3) Only high throughput supported interactions (HT). Both non-HT and HT were further subsampled to contain the same nodes, thus only differ in the number of edges. Non-HT edges were randomly removed until having the same number of edges as HT and the average clustering coefficient was stored; this was repeated 1000 times to create a null distribution. A Mann-Whitney U test was used to obtain a p-value of the high throughput average clustering coefficient subsample with respect to the non-HT sampled same sized distribution, and statistical significance was called at $α < 0.05$.

*Statistical analyses*

All other not previously mentioned analyses such as hypothesis tests, correlation coefficients estimation and so forth were performed using Python 2.7, using in-house written scripts and the *numpy*, *scipy* and *statmodels* modules.

**List of abbreviations**

1. GRN – Genetic regulatory network
2. ER – Erdos-Renyi
3. BA – Barabasi-Albert
4. HT – High-throughput curated graph subsample
5. Non-HT – Experimental curated (non-HT) graph subsample

*Acknowledgements*

This work was supported by the Programa de Apoyo a Proyectos de Investigación e Innovación Tecnológica (PAPIIT-UNAM) [IA200616 and IN205918 to JAFG]. We thank Marco Tello, Juan Escorcia, and Carlos Cruz for useful feedback and discussions. We thank Karla Vazquez-Prada and Brittany Mitchell for their help with manuscript proofreading. We also thank two anonymous reviewers for helpful suggestions.

*Author contributions*

JAFG conceived and supervised the study. JAFG and AICG designed the experiments. AICG performed the analyses. JAFG and AICG analyzed results and wrote the manuscript. All authors read and approved the final manuscript.

*Competing Interests*

The authors declare no competing interests.

*Data Availability*

The data analyzed during this study are available from Abasy Atlas at http://abasy.ccg.unam.mx. All data generated during this study are included in this published article (and its Supplementary Information files).

**Figures and Tables**

**Figure 1. Density and number of regulators exhibit trends with completeness in Abasy Atlas GRNs.**
a) Relationship between density values and number of genes in the network for all the existing GRNs in Abasy Atlas. b) Relationship between time, genomic coverage, and density of *E. coli* GRN. dsRNA-GRN including regulatory RNA interactions, strong-GRN with only strongly supported interactions as described in Regulon DB[21]. c) Relationship between the number of genes in a regulatory network and number of regulators.

**Figure 2. Goodness of fit of Abasy Atlas GRN P(k) to alternative probability distributions.**
a) Kolmogorov-Smirnov D (KS D) statistic of the GRN P(k) data against the MLE probability distributions. Higher values indicate a higher deviation (worse fit) from the fitted distribution. b) Log-likelihood ratio test score of power-law vs. other distributions. Higher values (red) indicate a preference for power law while smaller values (blue) indicate a preference for an alternative distribution (y axis labels). Blank spaces denote non-significant comparisons. All ER graphs initial parameters were generated by randomly sampling from the distribution of biologically equivalent measures (see Materials and methods). The scores depicted are the mean of 1000 random sampling experiments using a previously published information retrieval sampling scheme[32]. c,d) KS_D statistic assessing the goodness of fit of Erdos-Renyi graphs (sampled with the information retrieval scheme) to a Poisson (c) and Power-law P(k) distribution. As before higher values of KS D indicate a worse fit. Results represent the mean of 100 iterations of the sampling scheme for each combination of bait and coverage values. e) Heatmap depicting the goodness of fit differences for the same ER sampled networks, negative values would indicate a preference for power law whereas positive values indicate the expected preference for Poisson, all of this differences are statistically significant (see **Supplementary Fig. 3**). Detailed annotated subgraphs a and b are available in **Supplementary Fig. 4**.

**Figure 3. Property incompatibilities between GRN and theoretical network null models.**
a, b) BA and ER graphs were generated to span the range of densities observed in Abasy Atlas GRNs (No significant difference between the parametrized ER networks and the GRN distribution). c) Tail length distribution for the networks depicted in (a) (p < 0.001 Mann-Whitney U test). d) Distribution of average clustering coefficient for the networks depicted in (b) note the substantial differences between biological GRNs and both null models (p < 0.001 Mann-Whitney U test). e-f) Sampled ER network (previously reported as *power law*) properties (same networks as above) were calculated. Note that the better the fit to a *power law* (**Figure 2 c-e**) the higher the deviation of actual properties such as tail-length (e) and average clustering coefficient (f).

**Figure 4. GRN total number of interactions prediction.**
a, b and c) Models to estimate the total number of interactions in a GRN. a) Edge regression model (EdR). b) Density invariance model (DI) where *Dg* was obtained from average density of most complete graphs. c) Density proportionality model (DP), where density is modeled as an exponential decay. d, e, and f) Dependency between completeness comparison score and average clustering coefficient for the different models: Edge linear dependency (d) Density invariant (e) and the density proportionality factor (f). *E. coli* and *M. tuberculosis* GRNs are represented with different colors and markers. The comparison score enables a direct comparison of the GRN *completeness* as predicted by our interaction coverage (derived from the models) or the classical genomic coverage approach; it ranges from minus to positive infinity, with negative values indicating that the interaction coverage predicts the GRN to be less complete than the genomic coverage.

**Figure 5. *E. coli* purely high throughput subset of GRN contains an unexpectedly nonrandom low clustering coefficient.**
Average clustering coefficient distribution of the subsampled networks of *E. coli*[21]. All networks have the same number nodes. The number of edges in random-edge-removal networks and pure high throughput is the same. *M. tuberculosis* v2015 (83332_v2015_s15) complete GRN clustering coefficient is depicted in red for comparison.

Table 1. Characteristics and total number of interactions predicted by the density proportionality (DP), density Invariance (DI) or edge regress (EdR) approaches. *Estimate (95% confidence interval).

| | Actual number of interactions | Number of genes | Genomic coverage | Total interactions DP* | Density DP* | Total interactions DI* | Density DI* | Total interactions EdR* | Density EdR* |
|---|---|---|---|---|---|---|---|---|---|
| **196627_v2016_s17_eStrong** (*Corynebacterium glutamicum*) | 2911 | 3138 | 0.708413 | 7422 (4321-13450) | 0.00075 (0.00043-0.00136) | 8866 (4549-13182) | 0.00090 (0.00046-0.00133) | 7457 (6836-8054) | 0.00075 (0.00069-0.00081) |
| **224308_v2016_sSW16** (*Bacillus subtilis*) | 3040 | 4421 | 0.423886 | 11277 (6487-20734) | 0.00057 (0.00033-0.00106) | 17599 (9030-26168) | 0.00090 (0.00046-0.00133) | 10639 (9764-11481) | 0.00054 (0.00049-0.00058) |
| **451516_v2015_sRTB13** (*Staphylococcus aureus*) | 2039 | 2844 | 0.240155 | 6583 (3845-11879) | 0.00081 (0.00047-0.00146) | 7282 (3736-10827) | 0.00090 (0.00046-0.00133) | 6728 (6165-7269) | 0.00083 (0.00076-0.00089) |
| **511145_v2017_sRDB16_dsRNA** (*Escherichia coli*) | 6843 | 4497 | 0.537469 | 11514 (6619-21185) | 0.00056 (0.00032-0.00104) | 18210 (9343-27076) | 0.00090 (0.00046-0.00133) | 10827 (9937-11684) | 0.00053 (0.00049-0.00057) |
| **83332_v2015_s15** (*Mycobacterium tuberculosis*) | 6572 | 4091 | 0.62112 | 10259 (5917-18800) | 0.00061 (0.00035-0.00112) | 15070 (7732-22407) | 0.00090 (0.00046-0.00133) | 9820 (9011-10599) | 0.00058 (0.00053-0.00063) |

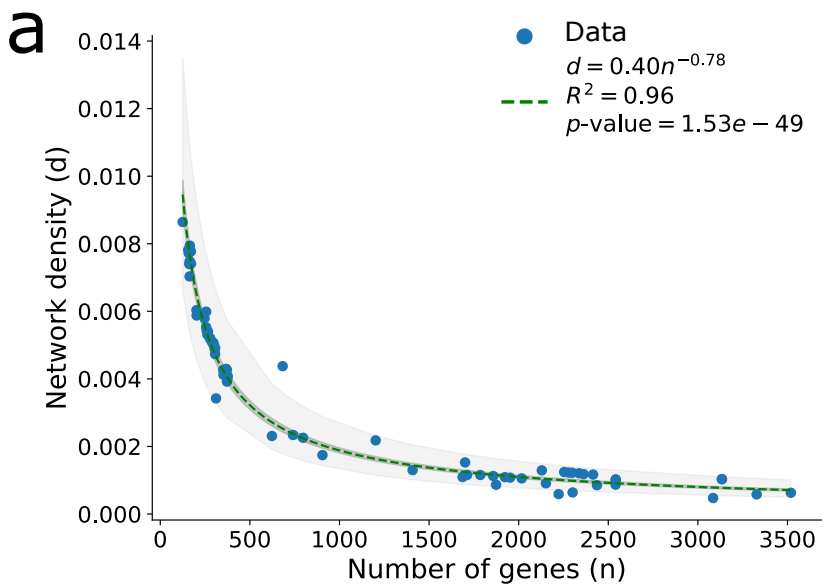

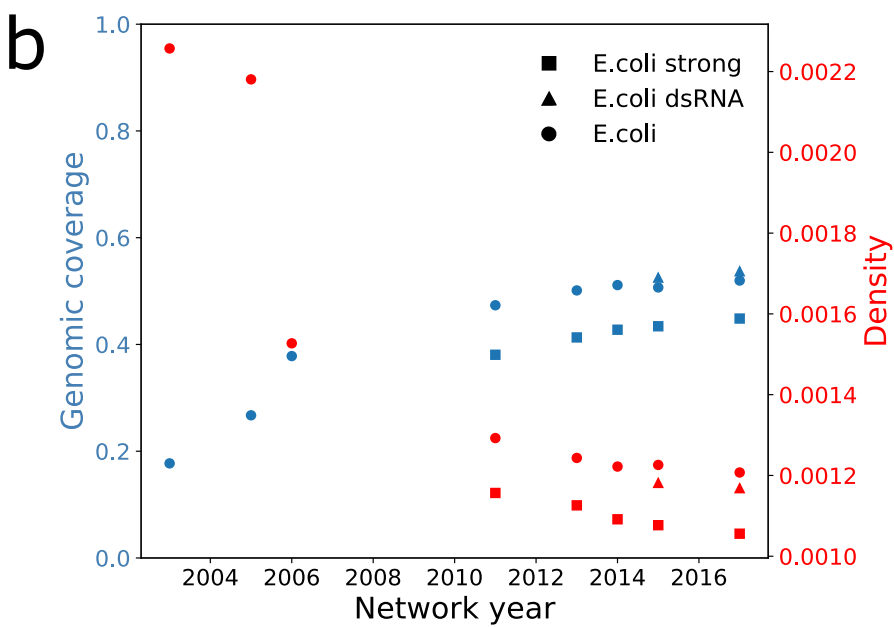

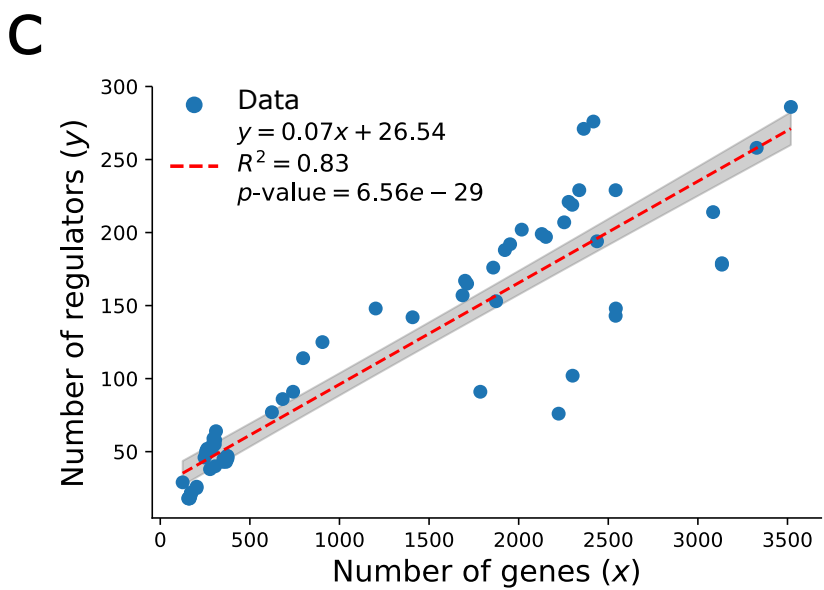

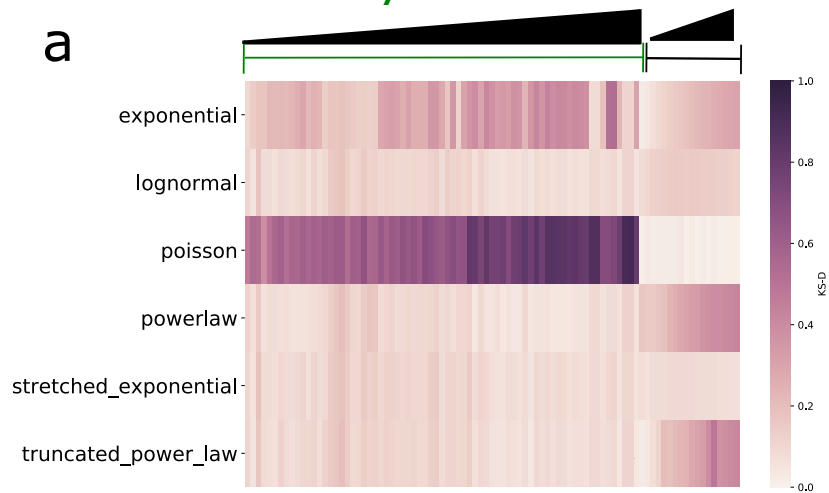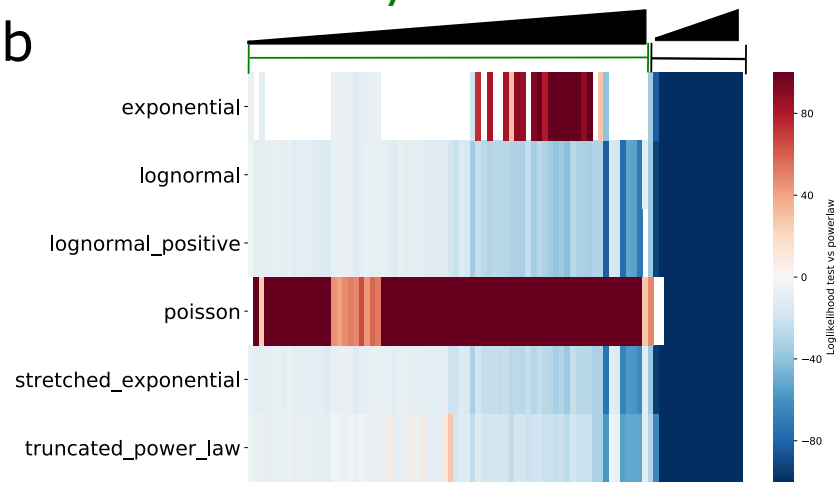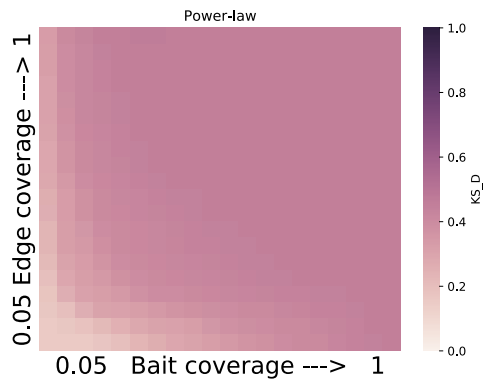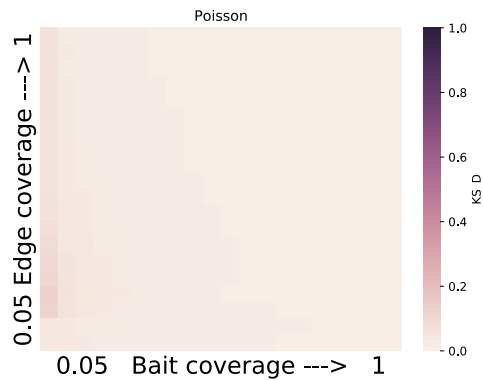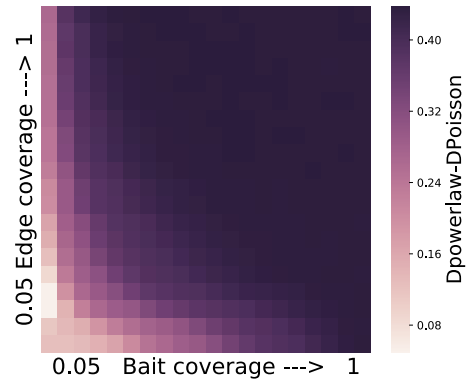

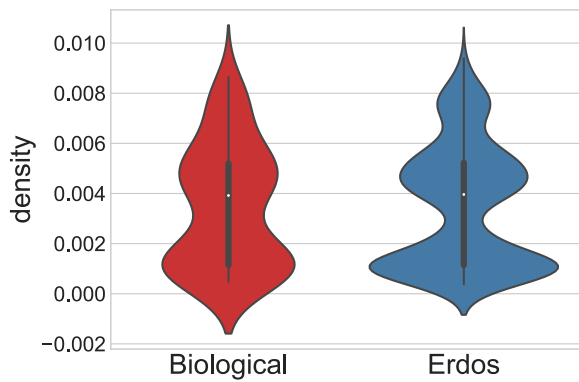
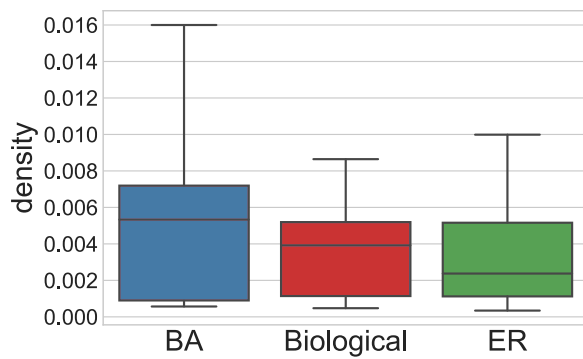
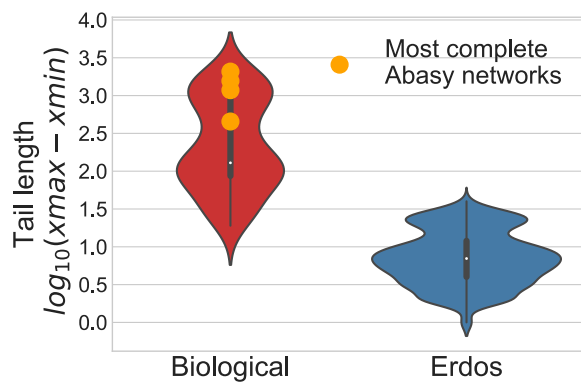
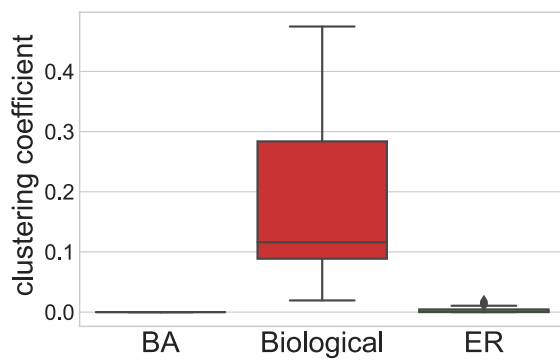
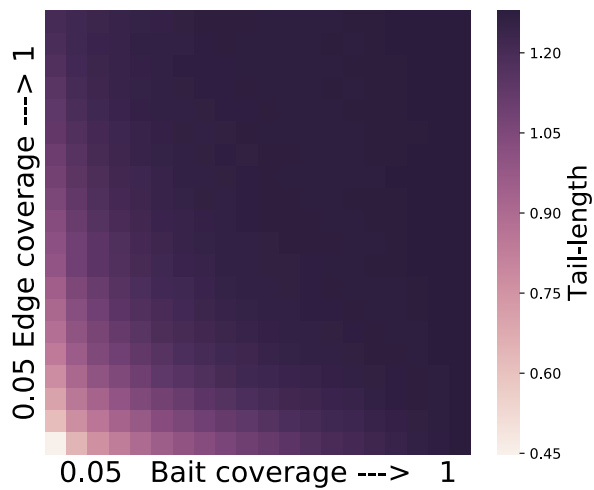
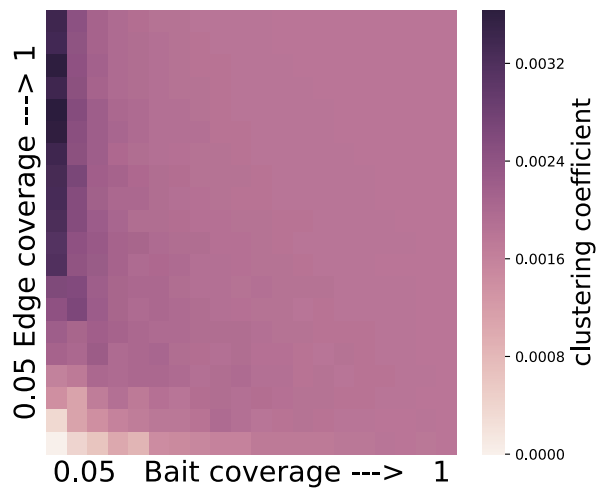

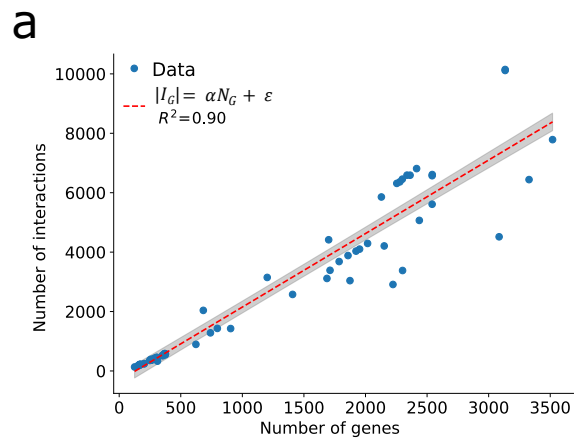
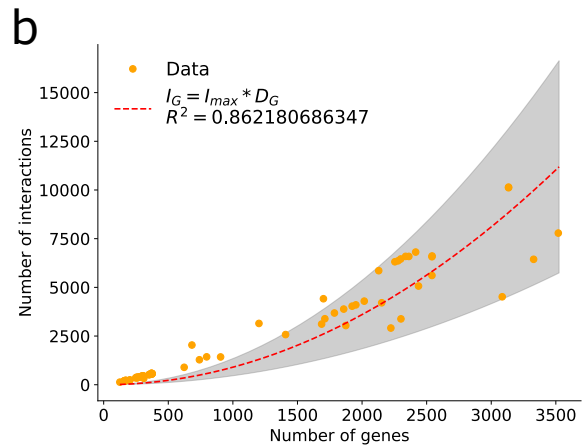
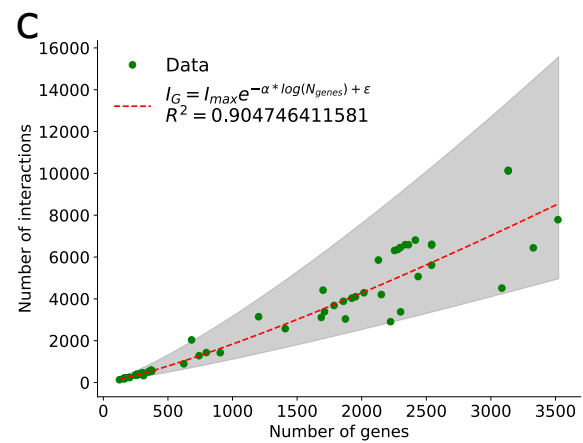
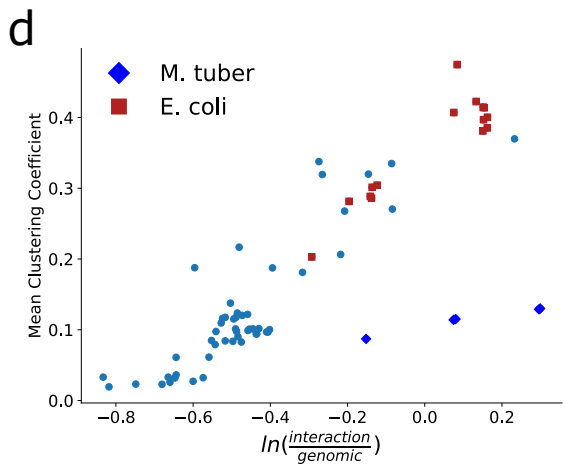
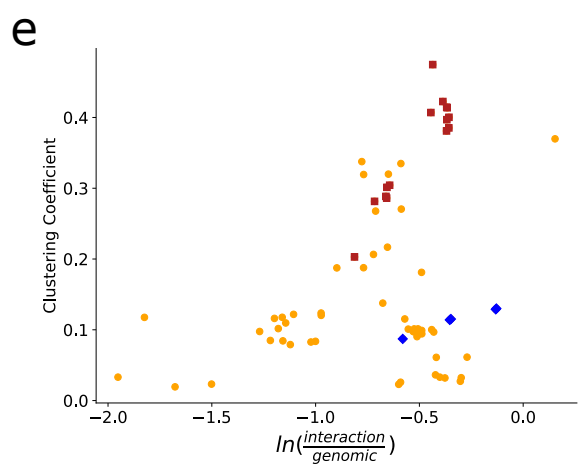
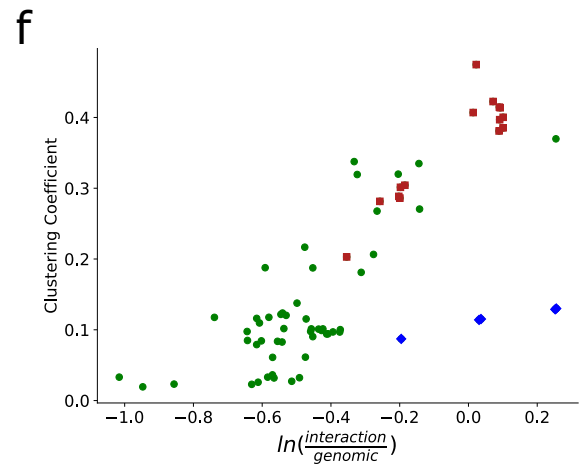

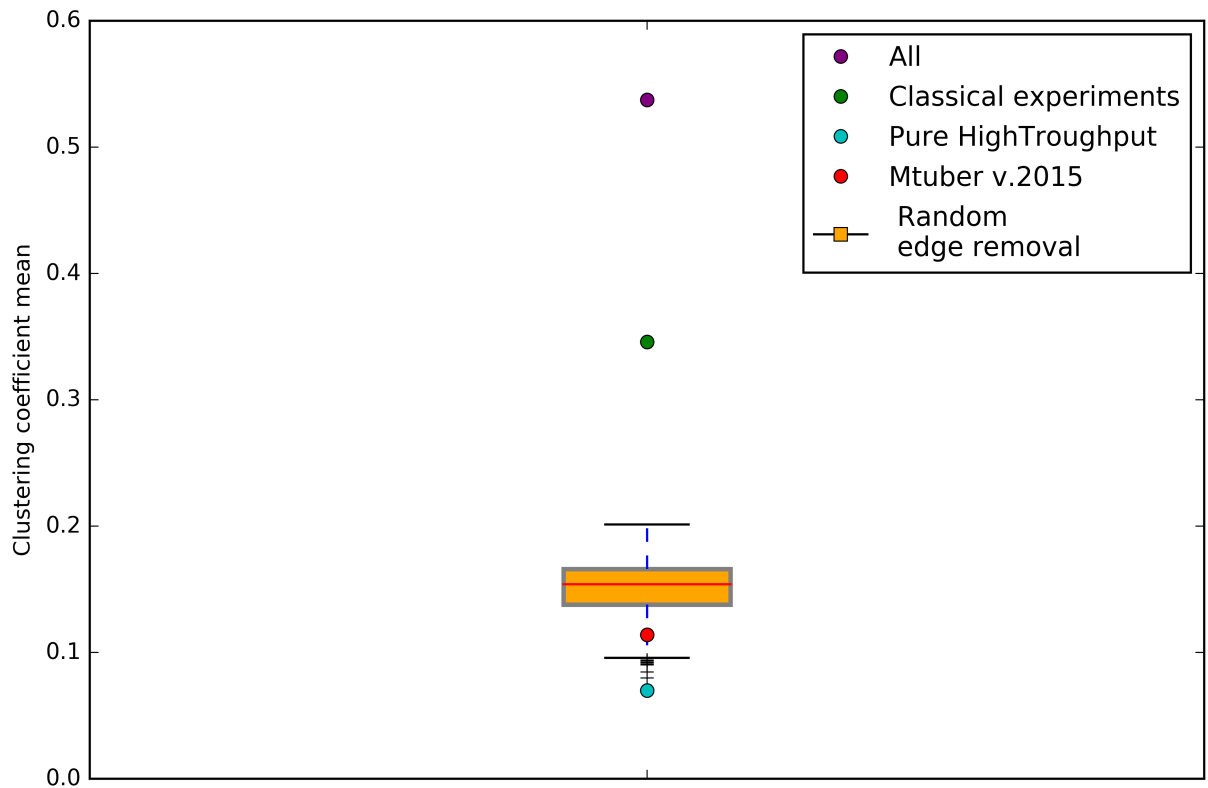

# Evolutionary constraints on the complexity of genetic regulatory networks allow predictions of the total number of genetic interactions

## Supplementary information

**Authors:** Adrian I. Campos-González[1,2] and Julio A. Freyre-González[1,]*

[1]Regulatory Systems Biology Research Group, Laboratory of Systems and Synthetic Biology, and [2]Undergraduate Program in Genomic Sciences, Center for Genomics Sciences, Universidad Nacional Autónoma De México, Av. Universidad s/n, Col. Chamilpa, 62210, Cuernavaca, Morelos, México.

**\*Corresponding author:** jfreyre@ccg.unam.mx

**Supplementary note 1: The May-Wigner stability theorem**

In the early 1970s, Gardner and Ashby empirically found that the stability of randomly connected large systems depends on their connectance[1]. They explored the system stability by modeling a set of nonlinear first-order differential equations whose coefficients, representing the interaction strengths among variables, were randomly obtained from a Gaussian distribution having zero mean and variance $α^2$. The change of the system state $\mathbf{x}(t)$ (where $\mathbf{x}(t) = (x_1(t), x_2(t), …, x_n(t))$) in time can be represented by the equation $d\mathbf{x}/dt = \mathbf{Ax}$, where $\mathbf{A}$ is the matrix of interaction strengths. The percentage of non-zero entries in this matrix was defined as the percentage of connectedness (connectance). Connectance is then analogous to density in graph theory as both quantify the fraction of existing interactions relative to the total possible.

Connectance (and consequently also density) has an important role in complexity theory as it quantifies the complexity of a system[1,2]. Robert M. May extended Gardner and Ashby's work to conclude that the stability of randomly connected systems depends on the number of variables ($n$), connectance ($C$), and interaction strength dispersion ($α^2$)[2]. The May-Wigner stability theorem says that randomly connected large systems are stable if $nC < 1/α^2$. Unstructured and structured systems have been shown to be bound by this theorem. An early theoretical work in structured model networks has suggested that such structures promote stability[3], and this was also observed for hierarchical networks under certain conditions[4]. However, a later theoretical study concluded that hierarchical and modular networks are less stable than random networks[5] and other showed that increasing modularity or the number of hierarchical layers tends to increase the probability of instability[6]. It has been theoretically suggested that optimizing multiple structural and dynamical constraints such as minimizing complexity and path length while increase robustness to dynamical perturbations will evolve modular scale-free networks[7].

**Supplementary note 2:** *Fitting xmin parameter for MLE of power-law distribution.*

The maximum likelihood estimate MLE for a power-law probability distribution has been derived[8,9] and is implemented in a python package called *powerlaw*[10]. Because of the parametrization of a power-law distribution, very small values' probability would tend towards infinity. The valid areas for which a power law would suffice a probability distribution are defined by a parameter called *xmin*. Although theoretically a valid approach for estimating the parameters for fitting power-law distributions, we identified an undesired behavior arising from using this *data trimming* parameter when comparing Erdos-Renyi (ER) and biological networks P(k). To explain the phenomena, and why setting the *xmin* parameter to 1 was proposed as a solution, we will discuss an example. Let us take *Escherichia coli* 2017 GRN including its RNA-mediated interactions (accession number: 511145_v2017_sRDB16_dsRNA), and generate an ER equivalent network having the same number of nodes and density. Notably ER graphs are constructed so that their node degree distribution follows a Poisson distribution. A histogram of the degrees of both networks would look like this:

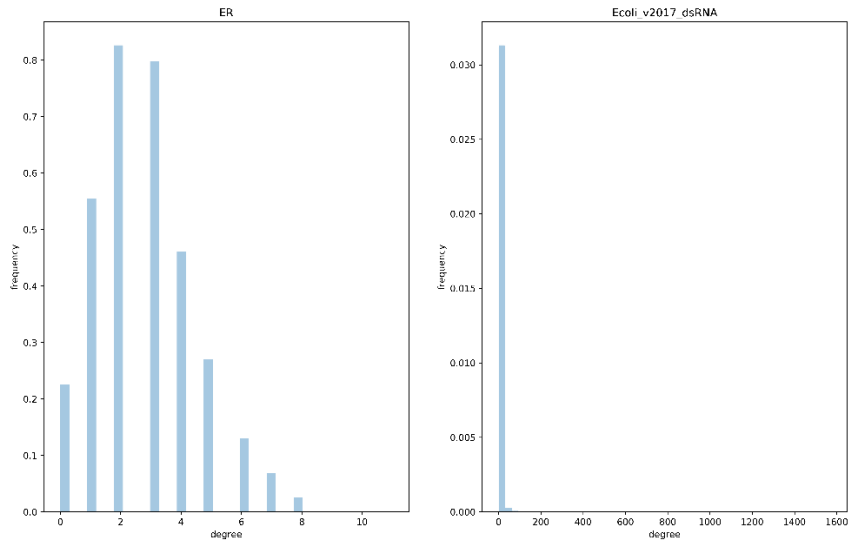

From this figure, we can already notice differences between the node degree distributions of these networks. First, while ER network seems normally distributed, *E. coli* GRN seems to follow a decaying trend and second, the biological network obviously follows a long-tailed distribution while the Poisson derived one does not.

If we were to fit a power-law distribution to the ER P(k) allowing for *xmin* to be a free parameter, most of the data of the ER network gets trimmed out. Importantly the trimmed data no longer seems to have a good fit to a poisson distribution (see below right panel).

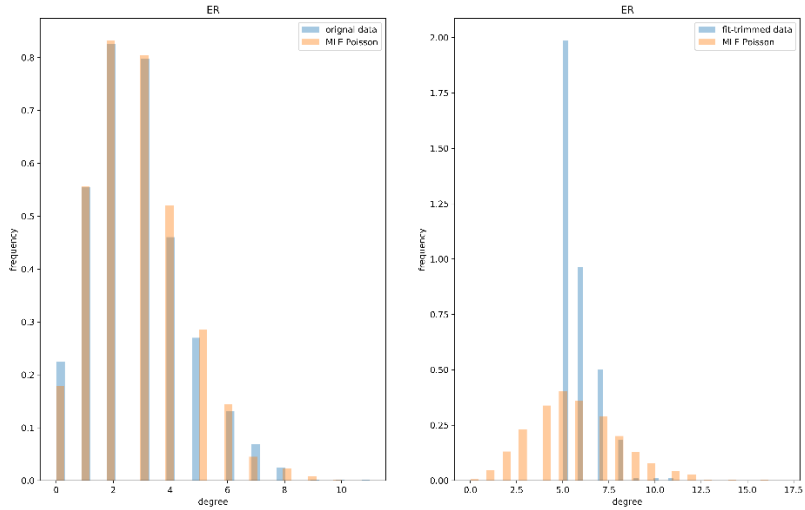

Allowing for a free *xmin* impedes us from understanding the true fit of the data to long-tailed distributions. Furthermore, a measure of the amount of information being ignored is not directly available from the estimates. Although in this example the *xmin* for ER and Biological networks are fairly close (5 and 3 respectively), their effects on trimming the data are far from being equal:

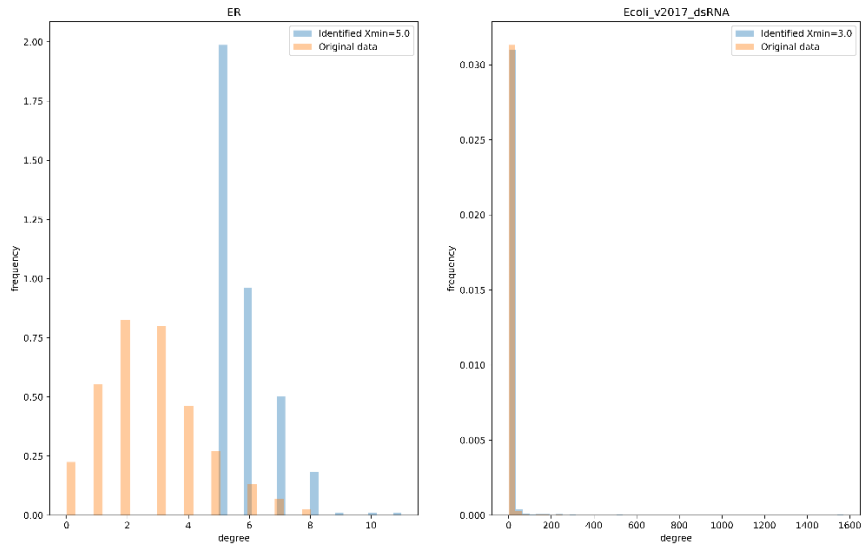

This result motivated us to repeat the analyses using a fixed *xmin* parameter of one, ensuring the use of all data available both for biological and theoretical networks.

**Supplementary figures**

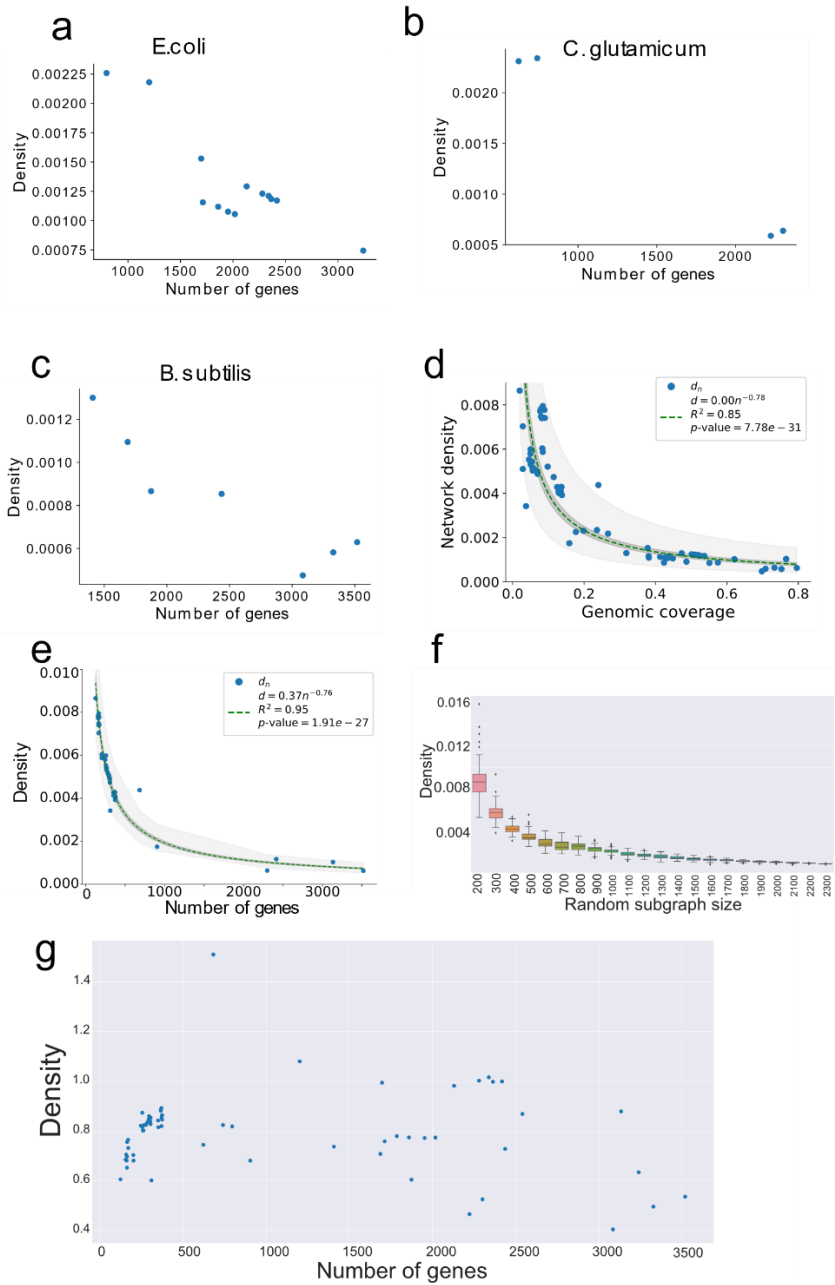

**Supplementary figure 1. Abasy GRNs present a trend towards a low density.**
a) Network density decreases with network completeness in historical reconstructions of *E. coli*. The same results are recapitulated in reconstructions of *C. glutamicum* (b) and *B. subtilis* (c). d) The use of genomic coverage as a completeness proxy does not affect the trend observed in network densities. e) The same density trend (as Fig. 1a) is observed when using a set of non-redundant GRNs. f) Random (snowball) sampling of *E. coli* 2013 network generates the same pattern as observed in (a), explaining the variability in networks with lower genomic coverage. g) Invariance observed in density when normalizing by randomly expected number of nodes in a GRN.

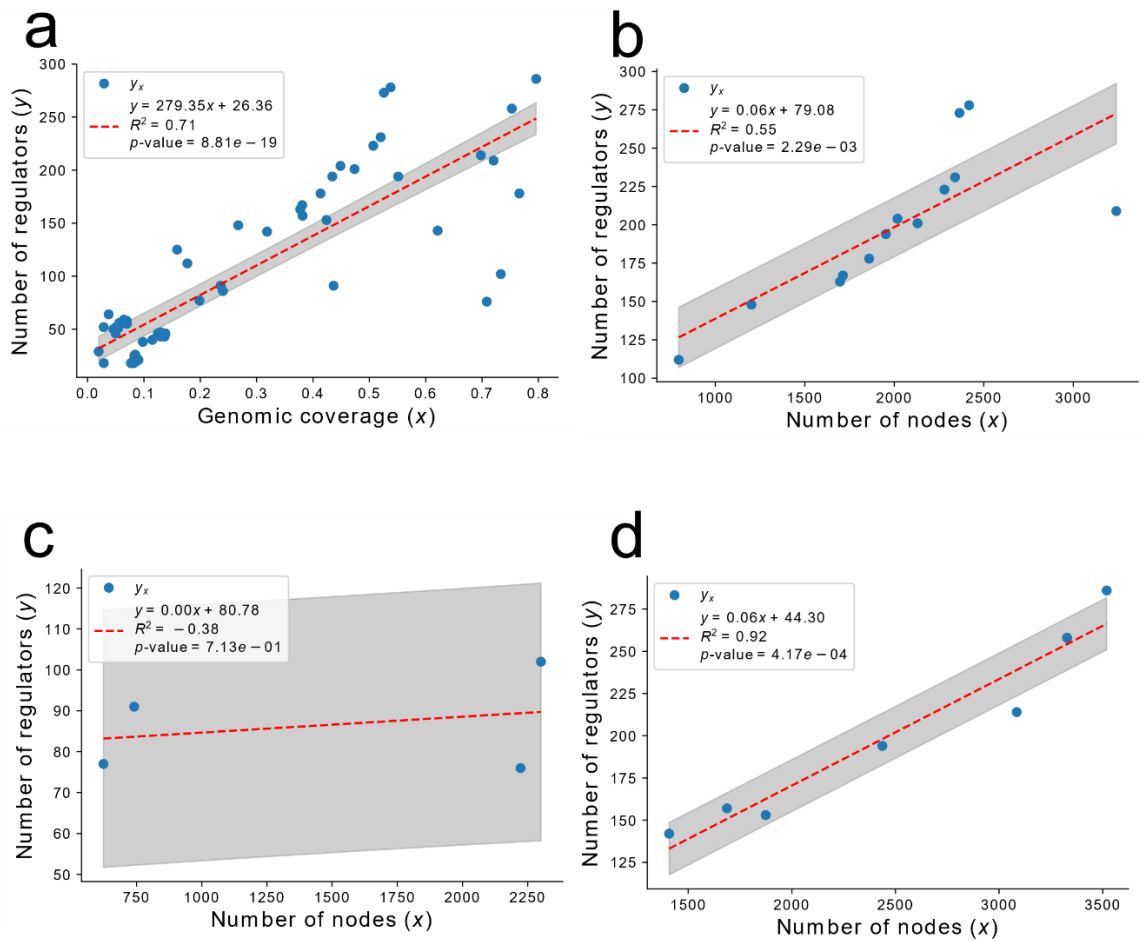

**Supplementary figure 2. Relationship between number of regulators and genes.**
The observed trend between completeness and number of regulators was recapitulated when using genomic coverage to assess completeness (a). The significant trend was also present in a set of historical reconstructions of *E. coli* (b), *C. glutamicum* (c) and independent *B. subtilis* (d) GRNs. We acknowledge a lack of power (data) for the historical reconstructions of *C. glutamicum* but included this result for reproducibility and openness.

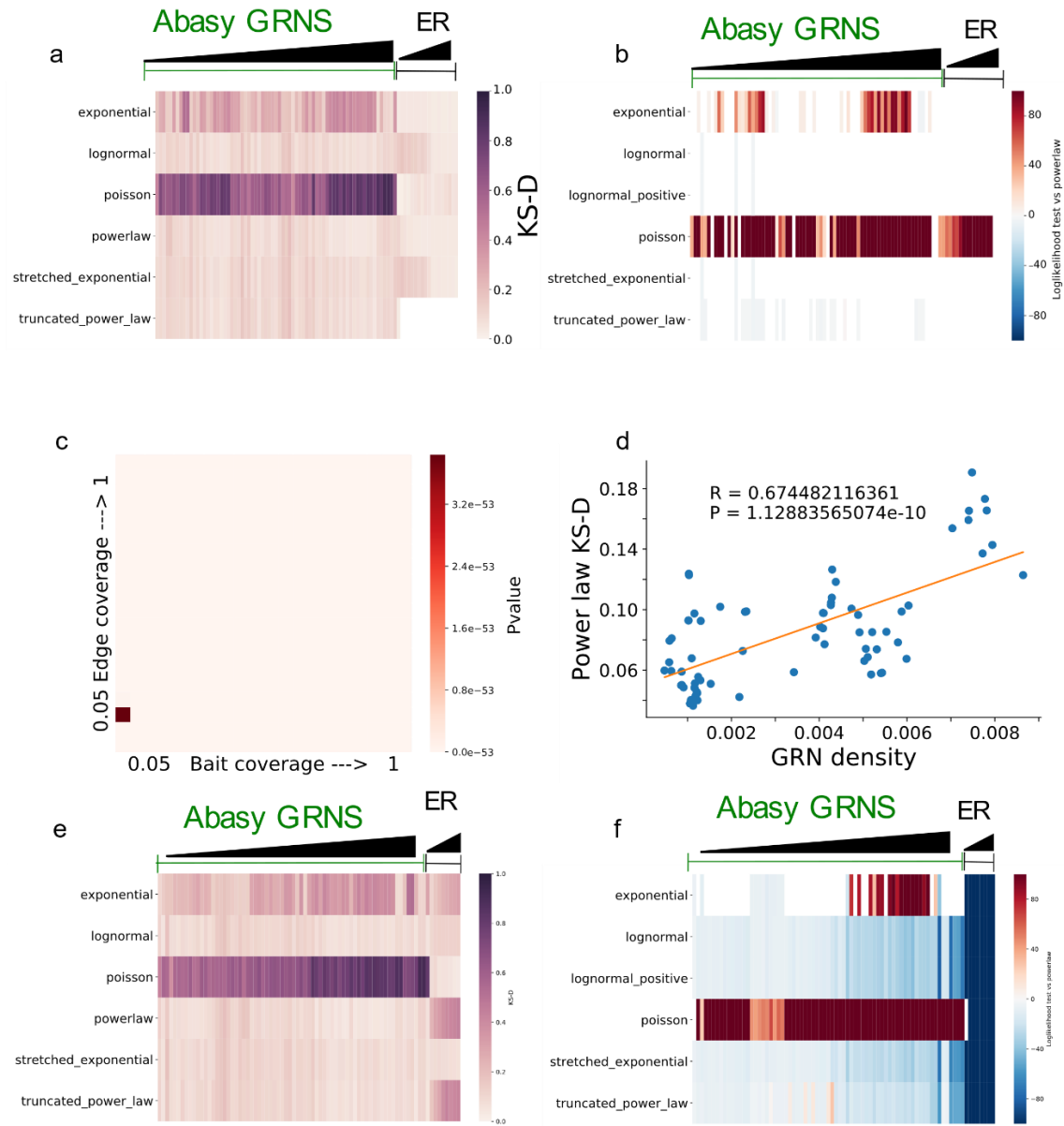

**Supplementary figure 3. GRN P(k) distribution follows a power-law distribution not likely arising from an artifice of sampling.**

a,b) Effect of the trimming parameter xmin when assessing goodness of fit of ER graphs to a set of probability distributions. These results correspond to the same approach as in **Figure 2**, but allowing xmin parameter to vary, thus not fitting all the connectivity data. Note that this causes **all** ER networks to have a good fit to and preference for a power law even when they are derived from a Poisson (see ER graphs and compare with **Figure 2**. c) P-values of the KS_D differences reported on **Figure 2e** note that all values are very significant. d) A negative correlation between goodness of fit to a power law and graph density was observed for the GRNs in Abasy. e,f) The results presented in Figure2 are recapitulated when using another sampling scheme for the ER networks: snowball sampling (**see methodology**).

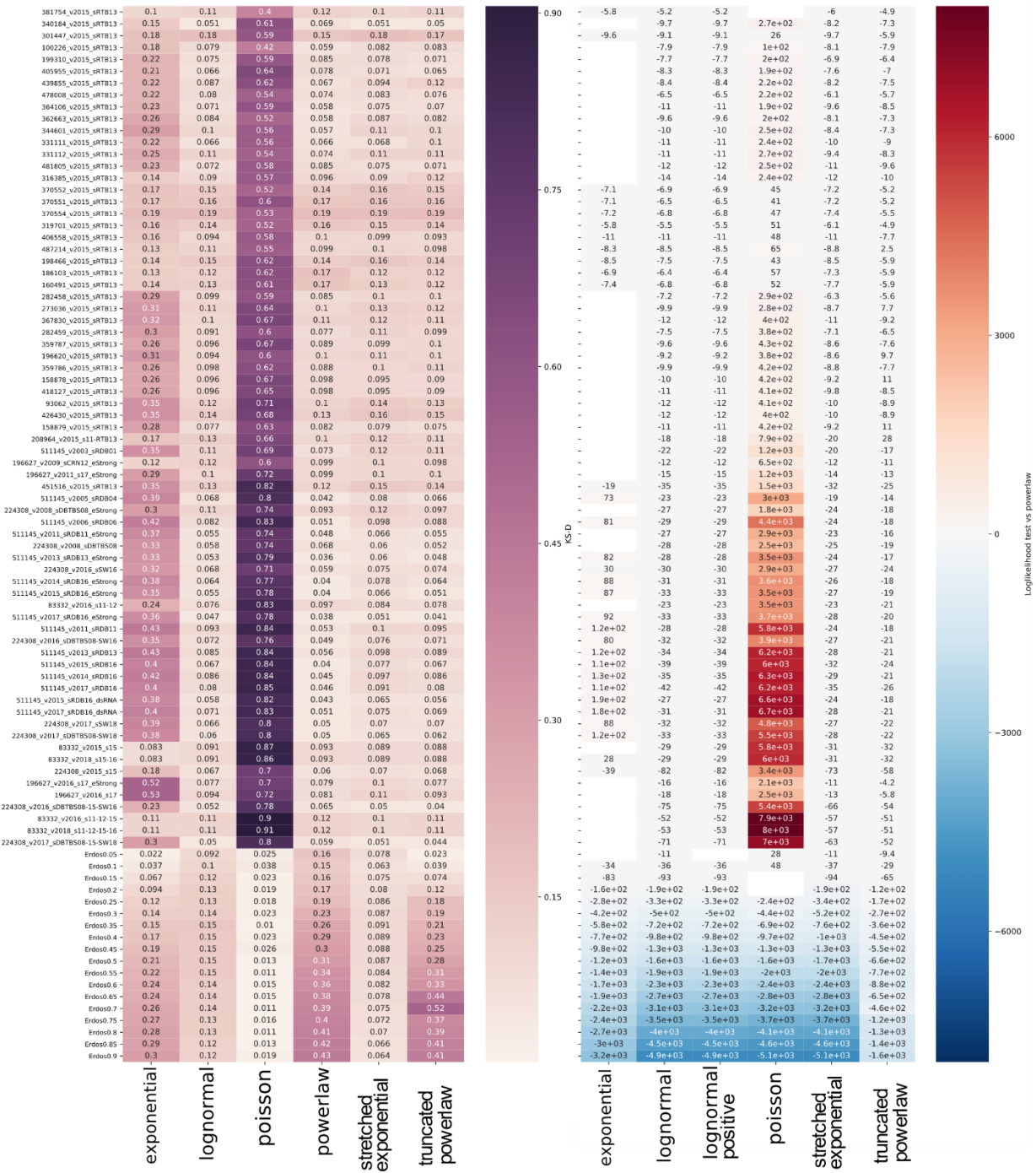

**Supplementary figure 4. Annotated version of Figure 2 from main manuscript.**
A detailed view with annotation scores is provided here.

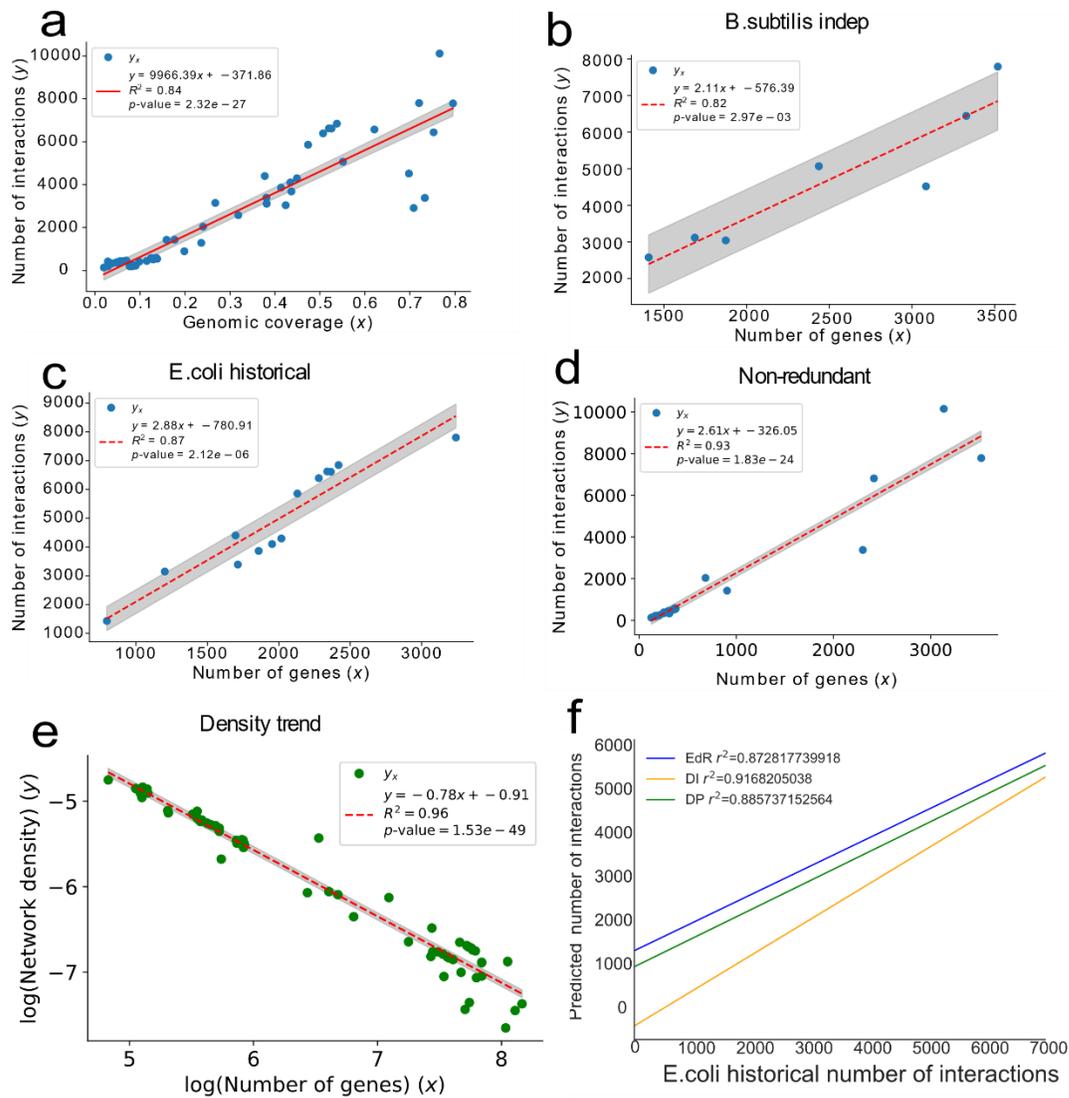

**Supplementary figure 5. Model generation and profiling.**
a) Relationship between genomic coverage and number of interactions. b and c) show the relationship between nodes and edges in historical reconstructions of *E. coli* and *B. subtilis*, respectively. d) Relationship between number of genes and number of interactions in the set of non-redundant networks. e) Density proportionality model parametrization. We model density as an exponential decay (by fitting a linear regression to the log transformed values) to predict number of nodes (**Fig 4c**). f) Comparison of the performance of the three models predicting *E. coli* historical reconstruction number of edges. Briefly, the three models were used to predict the number of interactions of the different *E. coli* GRNs and their accuracy was measured by the residual squared error ($R^2$)

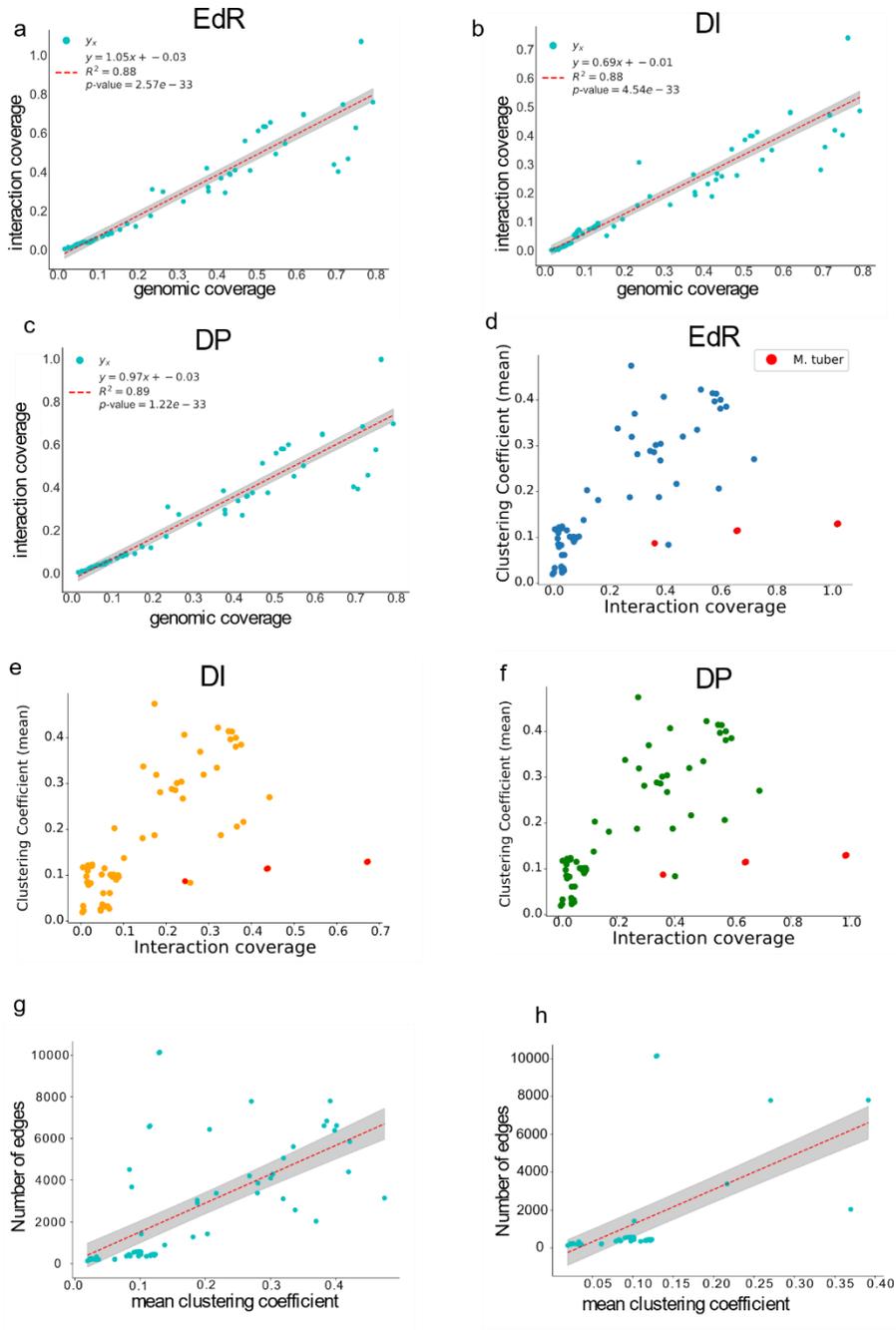

**Supplementary figure 6. Relationship between interaction coverage and clustering coefficient.**
a-c) A high correlation between the interaction coverage and genomic coverage using the edge regress, density invariant and density proportionality models respectively. d-f) Relationship between interaction coverage and mean network clustering coefficient using as estimator the models assuming EdR, density DI and DP models, respectively. g) Relationship between number of edges and clustering coefficients in Abasy GRNS. h) Same as in g but with the non-redundant networks.

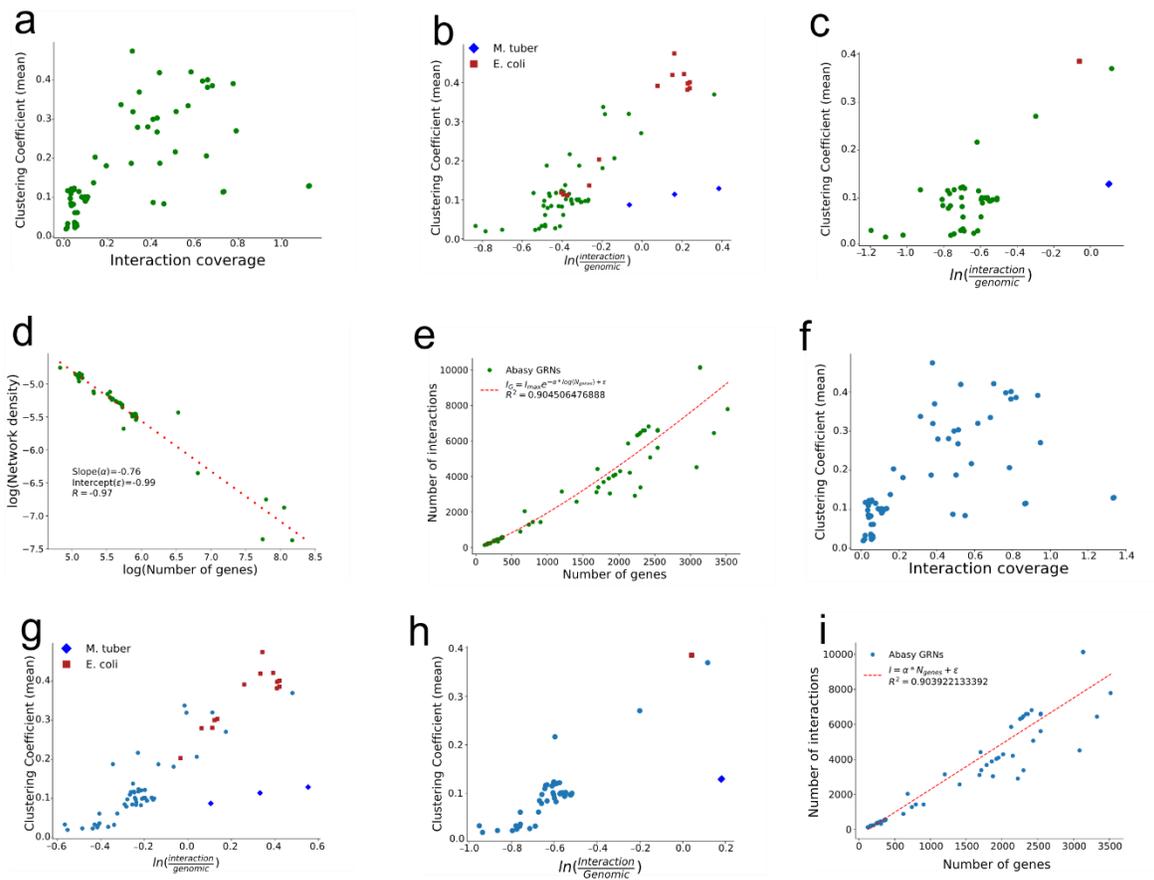

**Supplementary figure 7. Completeness estimation based on non-redundant networks.**
Model parameters were computed again using only a set of non-redundant, most complete networks (**see methods**). All panels are based on estimates using this set of networks. a-e) Model characteristics of density proportionality model. Panel b and c depict the same relationship, but on all Abasy GRNs (b) or only on the subset of non-redundant networks used to fit the models (c). d) Depicts the new parameters for modelling DP as an exponential decay to predict (e) number of interactions. f-i) Edge regress (EdR) model results when using only a set of non-redundant networks to parametrize the model. Note that overall results are similar to the ones presented in the main text using all networks to fit the models.